\documentclass[11pt,a4paper]{article}
\pdfoutput=1
\usepackage[utf8]{inputenc}
\usepackage{jheppub}

\usepackage{lmodern}

\usepackage{amsmath}
\usepackage{graphicx}
\usepackage{url}
\usepackage{epsfig}

\usepackage{multirow}

\newcommand{\squ}{{\tilde{q}}}

\newcommand{\glu}{{\tilde{g}}}
\newcommand{\sto}{{\tilde{t}}}

\newcommand{\alphas}{\alpha_{\mathrm{s}}}

\title{NNLL resummation for stop pair-production at the LHC}

\author[a]{Wim Beenakker,}
\author[b]{Christoph Borschensky,}
\author[b]{Raphael Heger,}
\author[c]{Michael Krämer,}
\author[b]{Anna Kulesza,}
\author[d]{Eric Laenen}

\affiliation[a]{Theoretical High Energy Physics, IMAPP, Faculty of Science, Mailbox 79, Radboud University Nijmegen, P.O. Box 9010, NL-6500 GL Nijmegen,\\
Institute of Physics, University of Amsterdam, Amsterdam, The Netherlands}
\affiliation[b]{Institute for Theoretical Physics, WWU Münster, D-48149 Münster, Germany}
\affiliation[c]{Institute for Theoretical Particle Physics and Cosmology, RWTH Aachen University, D-52056 Aachen, Germany}
\affiliation[d]{ITFA, University of Amsterdam, Science Park 904, 1018 XE, Amsterdam,\\
ITF, Utrecht University, Leuvenlaan 4, 3584 CE Utrecht,\\
Nikhef Theory Group, Science Park 105, 1098 XG Amsterdam, The Netherlands}

\emailAdd{w.beenakker@science.ru.nl}
\emailAdd{borschensky@uni-muenster.de}
\emailAdd{raphael.heger@uni-muenster.de}
\emailAdd{mkraemer@physik.rwth-aachen.de}
\emailAdd{anna.kulesza@uni-muenster.de}
\emailAdd{t45@nikhef.nl}

\abstract{The production of supersymmetric stop-antistop pairs at the Large Hadron Collider (LHC) is studied including corrections from soft-gluon resummation up to next-to-next-to-leading logarithmic (NNLL) accuracy in the Mellin-space approach. Additionally, corrections to the hard-matching coefficient at one-loop and Coulomb contributions at two-loop order are considered. The NNLL corrections enhance the cross section for all stop masses at centre-of-mass energies of 8 and 13 TeV compared to the previously calculated predictions at next-to-leading logarithmic (NLL) accuracy. Furthermore, a slight increase in the dependence on the additional stop-mixing parameters is observed.}


\begin{document}
\begin{flushright}
MS-TP-16-01\\TTK-16-01\\NIKHEF-2016-002\\
\end{flushright}
\maketitle

\renewcommand{\arraystretch}{1.2}

%%%%%%%%%%%%%%%%%%%%%%%%%%%%%%%%%%%%%%%%%%%
\section{Introduction}\label{s:intro}
%%%%%%%%%%%%%%%%%%%%%%%%%%%%%%%%%%%%%%%%%%%
The search for supersymmetry (SUSY) \cite{Wess:1973kz,Wess:1974tw,Fayet:1976et,Farrar:1978xj,Sohnius:1985qm,Martin:1997ns} is one of the main goals of the Large Hadron Collider (LHC). SUSY offers a rich variety of new particles and a number of dark matter candidates while at the same time providing a natural solution to the hierarchy problem and resulting in the unification of gauge couplings at high energy scales. For supersymmetry to be natural, it has to be realised close to the electroweak scale, in the reach of the current Run II of the LHC with a centre-of-mass energy of $\sqrt{S} = 13$ TeV. In particular, the partners of the coloured particles (squarks ($\squ$) and gluinos ($\glu$)) would be produced in copious amounts if their masses were around a few TeV. Experimental searches set a current lower mass limit of the coloured supersymmetric particles of around 1 TeV up to 1.8 TeV, depending on the specific SUSY model \cite{Aad:2015iea,Aad:2015baa,Khachatryan:2015lwa}.

The production of stops ($\sto$), the superpartners of the third generation top quarks, is an important special case, as, contrary to the case for light quarks, the top mass is very heavy and could lead to a strong mixing between the weakly interacting eigenstates of the stops and a large mass splitting between the physical stops \cite{Ellis:1983ed}. In a similar manner, the production of sbottoms, the superpartners of the bottom quarks, can be treated. Many SUSY scenarios assume the lighter stop mass eigenstate to be amongst the lightest particles in the spectrum, highlighting its special role in searches for supersymmetry at collider experiments \cite{Aad:2015pfx,Chatrchyan:2013xna,Chatrchyan:2013mya,Khachatryan:2015wza}.

In the framework of the Minimal Supersymmetric Standard Model (MSSM) \cite{Nilles:1983ge,Haber:1984rc} with R-parity conservation, supersymmetric particles are always produced in pairs through the collision of two hadrons $h_1$ and $h_2$. In the case of stops, the production proceeds through the following channels:
\begin{align*}
	h_1h_2 \to \sto_1\sto_1^* + X\quad\text{and}\quad\sto_2\sto_2^* + X.
\end{align*}
Here, $\sto_1$ and $\sto_2$ denote the lighter and the heavier mass eigenstate, respectively. The hadroproduction of mixed $\sto_1\sto_2^*$ or $\sto_2\sto_1^*$ is strongly suppressed, because these final states can only be realised at higher orders in the strong coupling $\alphas$ or through electroweak channels due to a strong suppression of top quarks in the initial state. The same holds for the production of equal-charge pairs like $\sto_1\sto_1$ or $\sto_2\sto_2$.

Accurate predictions for the production of stop pairs are important to derive mass exclusion limits, and, should supersymmetry be realised in nature, to precisely measure the masses and properties of the stops and other supersymmetric particles. Fixed-order cross sections for stop production have been calculated up to next-to-leading order (NLO) in $\alphas$ already some time ago \cite{Beenakker:1997ut}, with electroweak contributions having been computed \cite{Hollik:2007wf} and analysed \cite{Germer:2014jpa} more recently. An important part is played by the kinematical region where the production of stops proceeds close to the production threshold, i.e.\ where for the partonic centre-of-mass energy squared, $s \gtrsim 4m^2$ with $m$ being the mass of the produced stops. In particular for high stop masses, the NLO production cross section receives significant contributions from this region, where soft-gluon emission from the coloured initial- and final-state particles as well as gluon exchange between the slowly moving heavy final-state particles, the Coulomb contributions, lead to large corrections. Taking account of the soft-gluon corrections to all orders in the strong coupling can be achieved by means of threshold resummation techniques \cite{Sterman:1986aj,Catani:1989ne,Bonciani:1998vc,Contopanagos:1996nh,Kidonakis:1998bk,Kidonakis:1998nf}.

Corrections from soft gluons at next-to-leading logarithmic (NLL) accuracy have already been available for a while for all production processes of squarks and gluinos in the MSSM, including stops \cite{Kulesza:2008jb,Kulesza:2009kq,Beenakker:2009ha,Beenakker:2010nq,Beenakker:2011fu}. The same has been done within the framework of soft-collinear effective theory (SCET), also including the resummation of Coulomb contributions \cite{Beneke:2010da,Falgari:2012hx}. Approximate next-to-next-to-leading order (NNLO) corrections for stop-pair production, taking into account threshold-enhanced contributions at higher order, are also available, both from calculations in Mellin-moment space \cite{Langenfeld:2010vu} as well as employing the SCET formalism \cite{Broggio:2013uba}. Threshold resummation corrections up to next-to-next-to-leading logarithmic (NNLL) accuracy have been implemented for gluinos and first- and second-generation squarks in \cite{Beenakker:2011sf,Langenfeld:2012ti,Pfoh:2013edr,Beenakker:2013mva,Beneke:2013opa,Beneke:2014wda,Beenakker:2014sma}, and recently also for stops \cite{Broggio:2013cia}.

In this work, the impact of NNLL threshold resummation corrections on stop-antistop production is studied, updating the previous predictions for stop production including NLL corrections \cite{Beenakker:2010nq} and completing the analysis of \cite{Beenakker:2014sma} including all generations of squarks. The analysis is done in the Mellin-space approach, and for the LHC with centre-of-mass energies of $\sqrt{S} =$ 8 and 13 TeV with stop masses going up to 2.5 TeV and 3 TeV, respectively, to probe the resummation formalism in regions very close to threshold.

The structure of the paper is as follows. In section~\ref{s:theoryprod}, the properties of the partonic production processes of stop pairs are reviewed, and the colour-split Born cross sections are given, followed by an overview of the threshold resummation formalism in section~\ref{s:theoryresum}. Section~\ref{s:numerics} contains the numerical analysis, where the updated cross sections for stop-antistop production at an approximated NNLO including NNLL-resummed threshold corrections are presented, completed by a comparison to the similar light-flavoured squark-antisquark production as well as recent results using the SCET formalism. Section~\ref{s:conclusion} summarises the results in the conclusions.

%%%%%%%%%%%%%%%%%%%%%%%%%%%%%%%%%%%%%%%%%%%
\section{Stop pair-production}\label{s:theoryprod}
%%%%%%%%%%%%%%%%%%%%%%%%%%%%%%%%%%%%%%%%%%%
At the partonic level, stop pairs are dominantly produced through quark-antiquark annihilation and gluon fusion:
\begin{align}
	q\bar{q} &\to \sto_1\sto_1^*\quad\text{and}\quad\sto_2\sto_2^*,\notag\\
	gg &\to \sto_1\sto_1^*\quad\text{and}\quad\sto_2\sto_2^*.
\end{align}
Other channels are suppressed by higher orders of the QCD coupling $\alphas$. The mass eigenstates $\sto_1$ and $\sto_2$ originate from the mixing of the weak interaction eigenstates $\sto_L$ and $\sto_R$ as the proper superpartners of the left- and right-handed top quarks. For light-flavoured squarks, $\squ_1 \approx \squ_L$ and $\squ_2 \approx \squ_R$, as the mixing is proportional to the corresponding quark mass. Introducing a mixing angle $\theta_\sto$, the stop mass eigenstates can be written as $\sto_1 = \sto_L\cos\theta_\sto+\sto_R\sin\theta_\sto$ and $\sto_2 = -\sto_L\sin\theta_\sto+\sto_R\cos\theta_\sto$. The masses $m_{\sto_1}$ and $m_{\sto_2}$ are determined from diagonalising the stop mass matrix and depend on Standard Model and soft-supersymmetry-breaking parameters \cite{Ellis:1983ed}.

The corresponding cross sections at leading order, split-up corresponding to an SU(3) colour basis in the $s$-channel, read:
\begin{align}
	\sigma^{(0)}_{q_i\bar{q}_j\to\sto\sto^*,\mathbf{1}} &= 0,\label{qqbstst1}\\
	\sigma^{(0)}_{q_i\bar{q}_j\to\sto\sto^*,\mathbf{8}} &= \delta_{ij}\frac{\alphas^2\pi(N_c^2-1)}{12N_c^2s}\beta^3,\label{qqbstst8}\\
	\sigma^{(0)}_{gg\to\sto\sto^*,\mathbf{1}} &= \frac{\alphas^2\pi}{N_c(N_c^2-1)s}\left[\left(\frac{1}{2}+\frac{2m^2}{s}\right)\beta+\left(\frac{2m^2}{s}-\frac{4m^4}{s^2}\right)\ln\left(\frac{1-\beta}{1+\beta}\right)\right],\label{ggstst1}\\
	\sigma^{(0)}_{gg\to\sto\sto^*,\mathbf{8_A}} &= \frac{N_c^2}{2}\sigma^{(0)}_{gg\to\sto\sto^*,\mathbf{1}}+\frac{\alphas^2\pi N_c}{(N_c^2-1)s}\left[\left(-\frac{1}{6}+\frac{5m^2}{3s}\right)\beta+\frac{4m^4}{s^2}\ln\left(\frac{1-\beta}{1+\beta}\right)\right],\label{ggstst8A}\\
	\sigma^{(0)}_{gg\to\sto\sto^*,\mathbf{8_S}} &= \frac{N_c^2-4}{2}\sigma^{(0)}_{gg\to\sto\sto^*,\mathbf{1}},\label{ggstst8S}
\end{align}
where $N_c$ denotes the number of colours, $s$ the invariant partonic centre-of-mass energy squared, $m$ the mass of the produced stop, and $\beta = \sqrt{1-4m^2/s}$ the relative velocity between the two final-state particles. The colour indices $\mathbf{1}$, $\mathbf{8}$, $\mathbf{8_A}$, $\mathbf{8_S}$ refer to the singlet and (antisymmetric, symmetric) octet states in SU($N_c=3$), the value of $N_c$ will be, however, kept for a general SU($N_c$)-theory in all analytic expressions in this paper. The labels $i$, $j$ denote the initial-state quark flavours. The cross sections are valid both for the light as well as the heavy mass eigenstate, so in particular $m$ is the mass of the corresponding stop to be considered.

The production of stops is similar to light-flavoured squark-antisquark production ($\squ\squ^*$). For the gluon fusion channel, \eqref{ggstst1}--\eqref{ggstst8S} agree with the corresponding expressions for the light-flavoured squark-antisquark cross sections, excluding a factor of $2n_f$ with $n_f$ the number of light flavours, since only one type of stop is considered instead of a set of both the ``left- and right-handed''\footnote{Of course, chirality is not a property of scalar particles like the squarks, and this description only serves to distinguish between the partners of left- and right-handed quarks.} mass-degenerate light-flavoured squarks. The quark-antiquark channel, however, differs due to the fact that no top quarks are considered in the initial state. In comparison to the light-flavoured squark-antisquark production this leads to the absence of a Feynman diagram with gluino exchange, and in consequence to no colour-singlet contribution in this production channel. Furthermore, the colour-octet contribution is now $p$-wave suppressed, meaning that the stop pair has to be produced with a total angular momentum of at least $L = 1$ in order to balance out the spin from the gluon. This leads to a factor of $\beta^2$ near threshold, in addition to the phase-space suppression factor $\beta$.

Since QCD interactions are flavour-conserving at leading order, light-flavoured squarks are absent in Feynman diagrams. As a consequence of this and due to the above-mentioned lack of the gluino-exchange graph, only the mass of the produced stop explicitly appears in the cross sections \eqref{qqbstst1}--\eqref{ggstst8S}. At higher orders, however, interactions with gluinos and light-flavoured squarks arise in loops, leading to the appearance of additional parameters such as the gluino and light-flavoured squark masses $m_\glu$ and $m_\squ$, respectively, as well as the stop mixing parameter $\theta_\sto$ and the other stop mass, $m_{\sto_{2}}$ ($m_{\sto_{1}}$) in pair production of $\sto_1$ ($\sto_2$), respectively.

%%%%%%%%%%%%%%%%%%%%%%%%%%%%%%%%%%%%%%%%%%%
\section{NNLL resummation}\label{s:theoryresum}
%%%%%%%%%%%%%%%%%%%%%%%%%%%%%%%%%%%%%%%%%%%
The calculation of NNLL resummation is performed in an analogous manner to resummation for squark and gluino production \cite{Beenakker:2014sma}. However, for completeness, the formalism of threshold resummation will be briefly reviewed here. At higher orders in the perturbative series, terms of the form
\begin{align}
	\alphas^n\ln^k\beta^2,\quad k\leq 2n
\end{align}
appear in the cross sections. In the kinematical region close to the production threshold of the stops, $s \to (2m)^2$, these terms can become large and need to be included systematically up to all orders.

The inclusive hadronic cross section for the production of a stop-antistop pair can be written in its factorised form into long- and short-distance functions:
\begin{align}
	\sigma_{h_1h_2\to\sto\sto^*}(\rho,m^2) &= \sum_{i,j}\int dx_1dx_2d\hat{\rho}\,\delta\left(\hat{\rho}-\frac{\rho}{x_1x_2}\right)\notag\\
	&\qquad\qquad\qquad \times f_{i/h_1}(x_1,\mu^2)f_{j/h_2}(x_2,\mu^2)\sigma_{ij\to\sto\sto^*}(\hat{\rho},m^2,\mu^2),
\end{align}
where $\rho$ is the hadronic threshold variable, measuring the distance from the hadronic threshold, defined as
\begin{align*}
	\rho = \frac{4m^2}{S},
\end{align*}
$\sigma_{ij\to\sto\sto^*}(\hat{\rho},m^2,\mu^2)$ is the partonic cross section, $i$ and $j$ are the initial-state parton flavours, $f_{i/h_1}(x_1,\mu^2)$ and $f_{j/h_2}(x_2,\mu^2)$ are the parton distribution functions, $\mu$ is the common factorisation and renormalisation scale, and $x_1$ and $x_2$ are the momentum fractions of the partons inside the hadrons $h_1$ and $h_2$. Threshold resummation in this paper is carried out in Mellin-moment space, with the Mellin transform of the cross section defined as:
\begin{align}
	\tilde{\sigma}_{h_1h_2\to\sto\sto^*}(N,m^2) &\equiv \int_0^1 d\rho\,\rho^{N-1}\sigma_{h_1h_2\to\sto\sto^*}(\rho,m^2)\notag\\
	&= \sum_{i,j}\tilde{f}_{i/h_1}(N+1,\mu^2)\tilde{f}_{j/h_2}(N+1,\mu^2)\tilde{\sigma}_{ij\to\sto\sto^*}(N,m^2,\mu^2).
\end{align}
The logarithmically enhanced terms are now of the form $\alphas^n\ln^k N$, $k\leq 2n$, with the threshold limit $\beta\to 0$ corresponding to $N\to\infty$. The all-order summation of such logarithmic terms follows from near-threshold factorisation of the partonic cross section into functions describing the different kinds of gluon emission: hard, collinear, and wide-angle soft radiation \cite{Sterman:1986aj,Catani:1989ne,Bonciani:1998vc,Contopanagos:1996nh,Kidonakis:1998bk,Kidonakis:1998nf}. In terms of these functions, the partonic cross section can be written near threshold as:
\begin{align}
	&\tilde{\sigma}^{\mathrm{(res.)}}_{ij\to\sto\sto^*}(N,m^2,\mu^2) = \sum_I \tilde{\sigma}^{(0)}_{ij\to\sto\sto^*,I}(N,m^2,\mu^2)C_{ij\to\sto\sto^*,I}(N,m^2,\mu^2)\notag\\
	&\qquad\qquad\qquad\times\Delta_i(N+1,Q^2,\mu^2)\Delta_j(N+1,Q^2,\mu^2)\Delta^{\mathrm{(s)}}_{ij\to\sto\sto^*,I}(N+1,Q^2,\mu^2)\label{partresxsec}
\end{align}
with $I$ an index denoting the colour representation of the final state and $Q = 2m$ indicating the hard scale. The cross section $\tilde{\sigma}^{(0)}_{ij\to\sto\sto^*,I}$ is the Born cross section in Mellin-moment space projected onto the colour state $I$. The radiative factors
\begin{align}
	\ln\Delta_i(N,Q^2,\mu^2) &= \int_0^1 dz\frac{z^{N-1}-1}{1-z}\int_{\mu^2}^{Q^2(1-z)^2}\frac{dq^2}{q^2}A_i(\alphas(q^2))\\
	\ln\Delta^{\mathrm{(s)}}_{ij\to\sto\sto^*,I}(N,Q^2,\mu^2) &= \int_0^1 dz\frac{z^{N-1}-1}{1-z}\left[D_i(\alphas(Q^2(1-z)^2))+D_j(\alphas(Q^2(1-z)^2))\right.\notag\\
	&\qquad\qquad\qquad\qquad\left.+D_{ij\to\sto\sto^*,I}(\alphas(Q^2(1-z)^2))\right]
\end{align}
with each of the coefficients $A_i$, $D_i$, and $D_{ij\to\sto\sto^*,I}$ being a power series in $\alphas$, e.g.\ $A_i(\alphas) = \frac{\alphas}{\pi}A^{(1)}_i+\frac{\alphas^2}{\pi^2}A^{(2)}_i+\ldots$ (see e.g.\ \cite{Beenakker:2011sf} for the coefficients up to NNLL), are usually written in a form with the integrals evaluated:
\begin{align}
	\Delta_i\Delta_j\Delta^{\mathrm{(s)}}_{ij\to\sto\sto^*,I} = \exp\left[Lg_1(\alphas L)+g_2(\alphas L)+\alphas g_3(\alphas L)+\ldots\right]\label{resumlogs}.
\end{align}
These exponential functions resum the logarithms $L = \ln N$ originating from soft-collinear gluon radiation, with the function $g_1$ describing the leading logarithmic approximation (LL), whereas the inclusion of $g_2$ and $g_3$ leads to the NLL and NNLL approximation, respectively. Expressions for the functions $g_1$ and $g_2$ can be found in e.g.\ \cite{Kulesza:2009kq}, whereas $g_3$ is available in e.g.\ \cite{Beenakker:2011sf}. The matching coefficient $C_{ij\to\sto\sto^*,I}$ contains non-logarithmic terms as well as logarithmic terms of different origin in the Mellin moments. It has been shown that, close to threshold, these terms factorise into contributions from Coulomb exchange between final-state particles and a part containing off-shell hard contributions \cite{Beneke:2010da}:
\begin{align}
	C_{ij\to\sto\sto^*,I} = \left(1+\frac{\alphas}{\pi}\mathcal{C}^{\mathrm{Coul},(1)}_{ij\to\sto\sto^*,I}+\frac{\alphas^2}{\pi^2}\mathcal{C}^{\mathrm{Coul},(2)}_{ij\to\sto\sto^*,I}+\ldots\right)\left(1+\frac{\alphas}{\pi}\mathcal{C}^{(1)}_{ij\to\sto\sto^*,I}+\frac{\alphas^2}{\pi^2}\mathcal{C}^{(2)}_{ij\to\sto\sto^*,I}+\ldots\right).\label{matchingcoeffs}
\end{align}
Coulomb contributions originate from soft-gluon exchange between the two heavy final-state particles. They are enhanced near threshold due to their dependence on inverse powers of $\beta$. The expressions for the one- and two-loop Coulomb coefficients are given in \cite{Beenakker:2014sma}, see also \cite{Beneke:2013opa, Beneke:2014wda} for the additional spin-dependent contributions from non-Coulombic terms which contribute starting from two loops.

Close to threshold, the soft part in equation \eqref{partresxsec}, $\Delta_i\Delta_j\Delta^{\mathrm{(s)}}_{ij\to\sto\sto^*,I}$, factorises from a term often referred to as the hard function. In \cite{Broggio:2013uba}, this hard function has been calculated up to $\mathcal{O}(\alphas)$, split into colour channels. To identify the hard function at NLO with the one-loop hard-matching coefficient in the Mellin-space formalism $\mathcal{C}^{(1)}_{ij\to\sto\sto^*,I}$, the threshold behaviour of the NLO cross section \cite{Beneke:2009ye}\footnote{The given expansion only holds when stops are produced with zero angular momentum, i.e.\ in an $s$-wave colour channel. As discussed later, higher partial wave contributions for resummation at NNLL are not considered in the implementation presented in this paper. However, the NLO expansion for these terms is known, see e.g.\ \cite{Falgari:2012hx}.}
\begin{align}
	\sigma^{(1,\mathrm{thr})}_{ij\to\sto\sto^*,I} &= \sigma^{(0)}_{ij\to\sto\sto^*,I}\times\frac{\alphas}{4\pi}\Bigg\{-\frac{2\pi^2\kappa_{ij\to\sto\sto^*,I}}{\beta}+4(A^{(1)}_i+A^{(1)}_j)\left[\ln^2\left(\frac{8m\beta^2}{\mu}\right)+8-\frac{11\pi^2}{24}\right]\notag\\
	&\qquad\qquad+4(D^{(1)}_{ij\to\sto\sto^*,I}-4(A^{(1)}_i+A^{(1)}_j))\ln\left(\frac{8m\beta^2}{\mu}\right)-12D^{(1)}_{ij\to\sto\sto^*,I}+h^{(1)}_I+\mathcal{O}(\beta)\Bigg\}\label{thresnlo}
\end{align}
is compared to the NLO expansion of the Mellin-space resummation formula. In this equation, $h^{(1)}_I$ denotes the one-loop hard-matching coefficient in $\beta$ space. The value of $h^{(1)}_I$ can be found from the first-order coefficient of the NLO hard function $\mathbf{H}_{ij}$ calculated in \cite{Broggio:2013uba}, see also equation (26) therein. More precisely, numerical results for the first-order coefficient of the NLO hard function $\mathbf{H}_{ij}$ have been extracted from a computer code which is attached to the arXiv submission of \cite{Broggio:2013uba}. In order to obtain results for $h^{(1)}_I$, we have then numerically subtracted the one-loop Coulomb contributions from the first-order coefficient of the NLO hard function and subsequently taken the threshold limit $s \to (2m)^2$ in the outcome of this procedure. The results of this operation are then identified as $h^{(1)}_I$. The colour factor of the Coulomb contributions is defined as $\kappa_{ij\to\sto\sto^*,I} = C_2(I)/2-C_F$ with $C_2(I)$ being the quadratic Casimir invariant of the colour representation $I$. Taking the Mellin transform of equation \eqref{thresnlo} and comparing to the NLO expansion of equation \eqref{partresxsec}, the one-loop hard-matching coefficients in the Mellin-space formalism can be read off as:
\begin{align}
	\mathcal{C}^{(1)}_{ij\to\sto\sto^*,I} &= \frac{1}{4} h^{(1)}_I+\left(\frac{\pi^2}{24}+(\gamma_E -\ln 2)^2\right) (A^{(1)}_i+A^{(1)}_j)-D^{(1)}_{ij\to\sto\sto^*,I} (\gamma_E +1-\ln 2)\notag\\
	&\quad+\ln\left(\frac{\mu}{m}\right) \left((A^{(1)}_i+A^{(1)}_j) \ln\left(\frac{\mu}{m}\right)-2(\ln 2-\gamma_E) (A^{(1)}_i+A^{(1)}_j)-D^{(1)}_{ij\to\sto\sto^*,I}\right)
\end{align}
with $\gamma_E = 0.57721...$ the Euler-Mascheroni constant. It has been checked that the hard-matching coefficients defined in this way for the $gg$ initial state (as the only ones being relevant in this case, see the beginning of section~\ref{s:numerics}) yield the same numerical results as the ones for the similar process of light-flavoured squark-antisquark production, c.f.\ \cite{Beenakker:2013mva}, when choosing the stop masses the same as the common light-flavoured squark mass and setting the mixing angle $\theta_\sto$ to zero.

While the first and second order Coulomb contributions $\mathcal{C}^{\mathrm{Coul},(1)}_{ij\to\sto\sto^*,I}$ and $\mathcal{C}^{\mathrm{Coul},(2)}_{ij\to\sto\sto^*,I}$ are available, only the $\mathcal{O}(\alphas)$ hard-matching coefficient is known, so in the following, $\mathcal{C}^{(2)}_{ij\to\sto\sto^*,I}$ will be set to zero. In the numerical implementation of NNLL resummation, the expansion of \eqref{matchingcoeffs} up to $\mathcal{O}(\alphas^2)$ is used.

The proper inclusion of fixed-order corrections requires a matching procedure of the type:
\begin{align}
	&\sigma^{\mathrm{NNLO_{Approx}+NNLL}}_{h_1h_2\to\sto\sto^*}(\rho,m^2,\mu^2) = \sigma^{\mathrm{NNLO_{Approx}}}_{h_1h_2\to\sto\sto^*}(\rho,m^2,\mu^2)\notag\\
	&\qquad\qquad+\sum_{i,j}\int_{\mathrm{CT}}\frac{dN}{2\pi i}\rho^{-N}\tilde{f}_{i/h_1}(N+1,\mu^2)\tilde{f}_{j/h_2}(N+1,\mu^2)\notag\\
	&\qquad\qquad\times\left[\tilde{\sigma}^{\mathrm{(res,NNLL)}}_{ij\to\sto\sto^*}(N,m^2,\mu^2)-\tilde{\sigma}^{\mathrm{(res,NNLL)}}_{ij\to\sto\sto^*}(N,m^2,\mu^2)\Big|_{\mathrm{NNLO}}\right].\label{invmellin}
\end{align}
The approximate NNLO cross section is given as
\begin{align}
	\sigma^{\mathrm{NNLO_{Approx}}}_{h_1h_2\to\sto\sto^*}(\rho,m^2,\mu^2) = \sigma^{\mathrm{NLO}}_{h_1h_2\to\sto\sto^*}(\rho,m^2,\mu^2)+\Delta\sigma^{\mathrm{NNLO_{Approx}}}_{h_1h_2\to\sto\sto^*}(\rho,m^2,\mu^2),
\label{eq:nnloapprox}
\end{align}
where the first summand denotes the full NLO cross section calculation \cite{Beenakker:1997ut} and the second summand adds $\mathcal{O}(\alphas^2)$ terms that are enhanced near threshold in physical momentum space, i.e.\ terms that behave as inverse powers of $\beta$ and powers of logarithms of $\beta$, see equation (A.1) in \cite{Beneke:2009ye}. Furthermore, $\tilde{\sigma}^{\mathrm{(res,NNLL)}}_{ij\to\sto\sto^*}\Big|_{\mathrm{NNLO}}$ denotes the expansion of the resummed cross section up to NNLO in Mellin-moment space. The subtraction of this expansion from the resummed cross section $\tilde{\sigma}^{\mathrm{(res,NNLL)}}_{ij\to\sto\sto^*}$ is carried out in Mellin-moment space to avoid double counting of terms already taken into account in the fixed-order part. The difference between $\Delta\sigma^{\mathrm{NNLO_{Approx}}}_{h_1h_2\to\sto\sto^*}$ and the $\mathcal{O}(\alphas^2)$ terms of the expansion $\tilde{\sigma}^{\mathrm{(res,NNLL)}}_{ij\to\sto\sto^*}\Big|_{\mathrm{NNLO}}$ is of order $\mathcal{O}(1/N)$.

Lastly, the inverse Mellin transform is carried out with a contour CT according to the ``minimal prescription'' of reference \cite{Catani:1996yz} in order to obtain cross section predictions in physical momentum space.

%%%%%%%%%%%%%%%%%%%%%%%%%%%%%%%%%%%%%%%%
\section{Numerical results}\label{s:numerics}
%%%%%%%%%%%%%%%%%%%%%%%%%%%%%%%%%%%%%%%%
In this section, numerical results are presented for NNLL resummed cross sections matched to NNLO$_\mathrm{Approx}$ for the production of stop-antistop pairs at LHC energies of $\sqrt{S} =$~8,~13~TeV.

The QCD coupling $\alphas$ and the parton distribution functions at NLO and NNLO are defined in the $\overline{\mathrm{MS}}$ scheme with five active flavours, and a top-quark mass of $m_t = 173.21$ GeV is used \cite{Agashe:2014kda}. The renormalisation and factorisation scales are taken to be equal $\mu = \mu_R = \mu_F$. Unless stated otherwise, the common scale $\mu$ is taken to be the same as the light stop mass $m_{\sto_1}$.

The PDF sets chosen for the calculations are the MSTW2008 parton distributions \cite{Martin:2009iq}. For calculations at NNLL accuracy, MSTW2008 NNLO PDFs with a corresponding value of $\alphas(M_Z) = 0.117$, and for NLO and NLL results, MSTW2008 NLO parton distributions with a value of $\alphas(M_Z) = 0.120$ are used. The computer code \texttt{QCD-Pegasus} \cite{Vogt:2004ns} is used to derive the Mellin moments of the PDFs based on the MSTW parametrisation at the initial factorisation scale \cite{Martin:2009iq}. While there are updated PDF sets including LHC data available already since some time \cite{Ball:2014uwa,Harland-Lang:2014zoa,Dulat:2015mca} (see also \cite{Beenakker:2015rna}), the MSTW2008 set is used here to offer the possibility for an easy comparison to the recent study of squark and gluino production at NNLL \cite{Beenakker:2014sma}. A phenomenological study of stop-antistop production including the effects of different PDF sets is forthcoming.

The light-flavoured squark and gluino masses and the additional stop mixing parameters are set according to benchmark point 40.2.5 of \cite{AbdusSalam:2011fc}. For the predictions presented in this paper, the values from table~\ref{tbl:benchmark}~(a) are used. The predictions from \cite{Broggio:2013cia} are also based on benchmark point 40.2.5, however with slightly different numerical values. In comparisons to these results, their parameters have been used in the cross section calculations, as shown in table~\ref{tbl:benchmark}~(b). To study the mass dependence of the results, the stop mass $m_{\sto_1}$ is varied while keeping all other SUSY parameters fixed. Due to the additional parameters first appearing at NLO, the dependence of the cross section on these parameters is small, as argued in \cite{Beenakker:1997ut} (see also \cite{Beenakker:2010nq}).
\begin{table}
	\centering
	\begin{tabular}{|c||c|c|}
		\hline
		Benchmark point 40.2.5 & (a) & (b)\\\hline\hline
		$m_{\sto_1}$ [GeV] & 1085 & 1087.15\\
		$m_{\sto_2}$ [GeV] & 1320.85 & 1319.87\\
		$m_{\glu}$ [GeV] & 1492.57 & 1489.98\\
		$m_{\squ}$ [GeV] & 1496 & 1460.3\\
		$\sin 2\theta_\sto$ & 0.669 & 0.685\\\hline
	\end{tabular}
	\caption{SUSY parameters used in the calculation of stop cross sections for benchmark point 40.2.5 of \cite{AbdusSalam:2011fc}. The numerical values of (a) are used in this work (``default values''), while the ones from (b) were used in \cite{Broggio:2013cia}.}
	\label{tbl:benchmark}
\end{table}

In the plots and tables, the following labels will be used to denote the accuracy of the cross sections:
\begin{itemize}
	\item LO and NLO, denoting the leading- and next-to-leading order contributions \cite{Beenakker:1997ut},
	\item NLO+NLL, denoting the resummed soft-gluon corrections at NLL matched to NLO as calculated in \cite{Beenakker:2010nq},
	\item NLO+NNLL$_{h+\mathrm{Coul}(1)}$, denoting the resummed NNLL corrections matched to NLO, including the matching coefficient in \eqref{matchingcoeffs} up to $\mathcal{O}(\alphas)$, i.e.\ $C_{ij\to\sto\sto^*,I}\big|_{\mathrm{NLO}} = 1+\frac{\alphas}{\pi}\left(\mathcal{C}^{\mathrm{Coul},(1)}_{ij\to\sto\sto^*,I}+\mathcal{C}^{(1)}_{ij\to\sto\sto^*,I}\right)$, as a comparison to the results given in \cite{Broggio:2013cia} (using parameters according to table~\ref{tbl:benchmark}~(b)),
	\item NLO+NNLL, denoting the resummed NNLL corrections matched to NLO, including hard-matching coefficients up to $\mathcal{O}(\alphas)$ and Coulomb contributions up to $\mathcal{O}(\alphas^2)$, $C_{ij\to\sto\sto^*,I}\big|_{\mathrm{NNLO}} = 1+\frac{\alphas}{\pi}\left(\mathcal{C}^{\mathrm{Coul},(1)}_{ij\to\sto\sto^*,I}+\mathcal{C}^{(1)}_{ij\to\sto\sto^*,I}\right)+\frac{\alphas^2}{\pi^2}\left(\mathcal{C}^{\mathrm{Coul},(1)}_{ij\to\sto\sto^*,I}\mathcal{C}^{(1)}_{ij\to\sto\sto^*,I}+\mathcal{C}^{\mathrm{Coul},(2)}_{ij\to\sto\sto^*,I}\right)$. This allows for an investigation of the impact of the $\mathcal{O}(\alphas^2)$ terms in the matching coefficient, i.e.\ the two-loop Coulomb contribution as well as the cross-term consisting of the one-loop Coulomb and hard-matching coefficients, as compared to NLO+NNLL$_{h+\mathrm{Coul}(1)}$,
	\item NNLO$_{\mathrm{Approx}}$, as given by Eq.~\ref{eq:nnloapprox}, denoting the approximated NNLO contributions in the threshold limit from \cite{Beneke:2009ye} added to the full NLO result,
	\item NNLO$_{\mathrm{Approx}}$+NNLL, denoting the resummed NNLL corrections matched to the approximated NNLO result, including hard-matching coefficients up to $\mathcal{O}(\alphas)$ and Coulomb contributions up to $\mathcal{O}(\alphas^2)$. The matching coefficients $C_{ij\to\sto\sto^*,I}$ are taken as $C_{ij\to\sto\sto^*,I}\big|_{\mathrm{NNLO}}$ given above. This prediction differs from NLO+NNLL by terms of $\mathcal{O}(1/N)$, i.e.\ power-suppressed in $N$. Unless stated otherwise, the values of MSSM parameters listed in table~\ref{tbl:benchmark}~(a) are used. This is the default prediction presented in this paper.
\end{itemize}
It should be noted that colour channels for which the stops have to be produced with an angular momentum larger than zero, the so-called $p$-wave channels, are suppressed by an additional factor of $\beta^2$ close to threshold. In the case considered here, the $p$-wave channels are the octet channel $\mathbf{8}$ for the $q\bar{q}$ initial state and the antisymmetric octet channel $\mathbf{8_A}$ of the $gg$ initial state. Since for $p$-wave channels, the full factorisation of soft and hard parts has not completely been proven at NNLL accuracy yet\footnote{See \cite{Falgari:2012hx} for a related discussion.}, their contributions are calculated and taken into account only up to NLO+NLL accuracy\footnote{It has been checked that the terms of NNLO$_{\mathrm{Approx}}$+NNLL accuracy in the $p$-wave channels modify the total cross section by only a few permille, and thus are well within the theoretical uncertainty given in Section 4. For a consistent treatment of the PDFs, the $p$-wave contributions at NLO+NLL accuracy have been computed with NNLO PDFs, as they are added to the NNLO$_{\mathrm{Approx}}$+NNLL $s$-wave channels. It has been checked that the numerical impact of choosing NLO or NNLO PDFs for the $p$-wave channels is negligible.}. This means that for the two channels mentioned above, the NNLL function $g_3$ of equation~\eqref{resumlogs} is not taken into account, the matching coefficients of equation~\eqref{matchingcoeffs} are set to 1, $C_{q\bar{q}\to\sto\sto^*,\mathbf{8}} = C_{gg\to\sto\sto^*,\mathbf{8_A}} = 1$, and the matching of equation~\eqref{invmellin} is performed at NLO, with the additional NNLO$_\mathrm{Approx}$ corrections being set to zero, $\Delta\sigma^{\mathrm{NNLO_{Approx}}}_{q\bar{q}\to\sto\sto^*,\mathbf{8}} = \Delta\sigma^{\mathrm{NNLO_{Approx}}}_{gg\to\sto\sto^*,\mathbf{8_A}} = 0$.

While the publicly available code \texttt{Prospino} \cite{prospino} has been used for the numerical calculations of the NLO cross section, the resummed contributions have been obtained and checked using two independent computer codes.

\subsection{Threshold behaviour of the partonic cross section}
Scaling functions are dimensionless functions that show the behaviour of the cross section at a partonic level, not being influenced by the convolution with the PDFs. They serve as a useful test to compare the full partonic cross section to various approximations that are only valid in certain kinematical regions. The scaling functions can be extracted from the partonic cross section \cite{Beenakker:1997ut}:
\begin{align}
	\hat{\sigma}_{ij} = \frac{\alphas^2}{m_{\sto_1}^2}\left\{f^B_{ij}+4\pi\alphas\left[f_{ij}+\bar{f}_{ij}\ln\left(\frac{\mu^2}{m_{\sto_1}^2}\right)\right]\right\},
\end{align}
where $f^B_{ij}$ is the Born contribution and $f_{ij}$ and $\bar{f}_{ij}$ are the NLO scaling functions.

\begin{figure}[t]
	\centering
	\includegraphics[width=0.7\columnwidth]{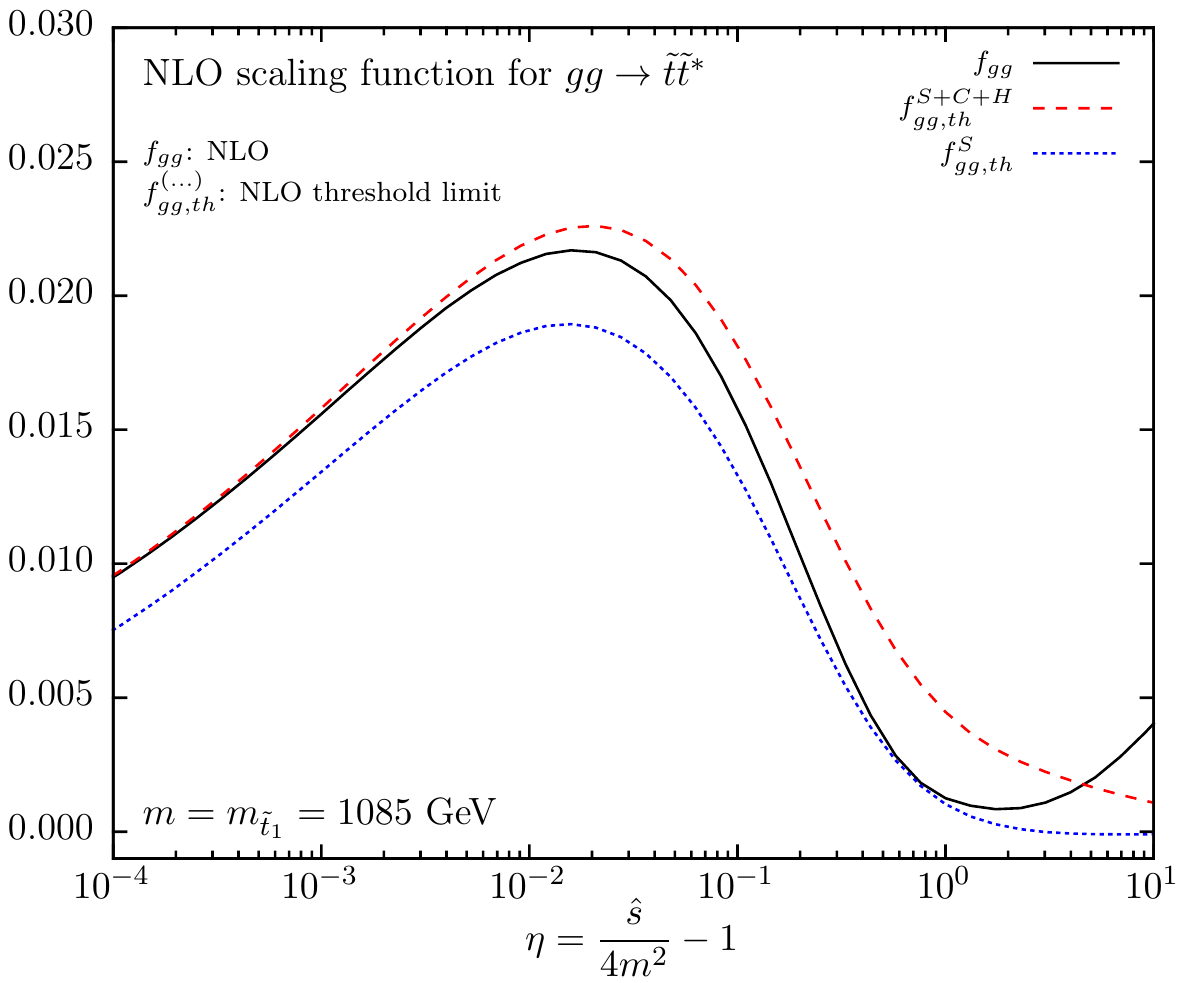}
	\caption{NLO scaling function for the $gg$ initial state for different threshold approximations and the full \texttt{Prospino} result.}
	\label{fig:st_scaling}
\end{figure}

Figure~\ref{fig:st_scaling} shows the NLO scaling function $f_{gg}$ for the $gg$ initial state for various approximations and the full NLO result extracted from \texttt{Prospino}. While the threshold approximation of the scaling function $f_{gg,th}$ does not agree very well with the full result $f_{gg}$ when only taking into account soft contributions ($S$), its behaviour significantly improves close to threshold ($\eta\to 0$) if one also adds the one-loop Coulomb and hard-matching coefficients ($S+C+H$). Especially the Coulomb term brings the approximation closer to the full result in the threshold region. Therefore, it is very important to incorporate the information contained in the one-loop hard matching coefficient and the Coulomb corrections into the full resummed expression, as done at the NNLL accuracy.

\subsection{Predictions for stop-antistop production at the LHC}
\begin{figure}[t]
	\centering
	\begin{tabular}{cc}
		\includegraphics[width=0.47\columnwidth]{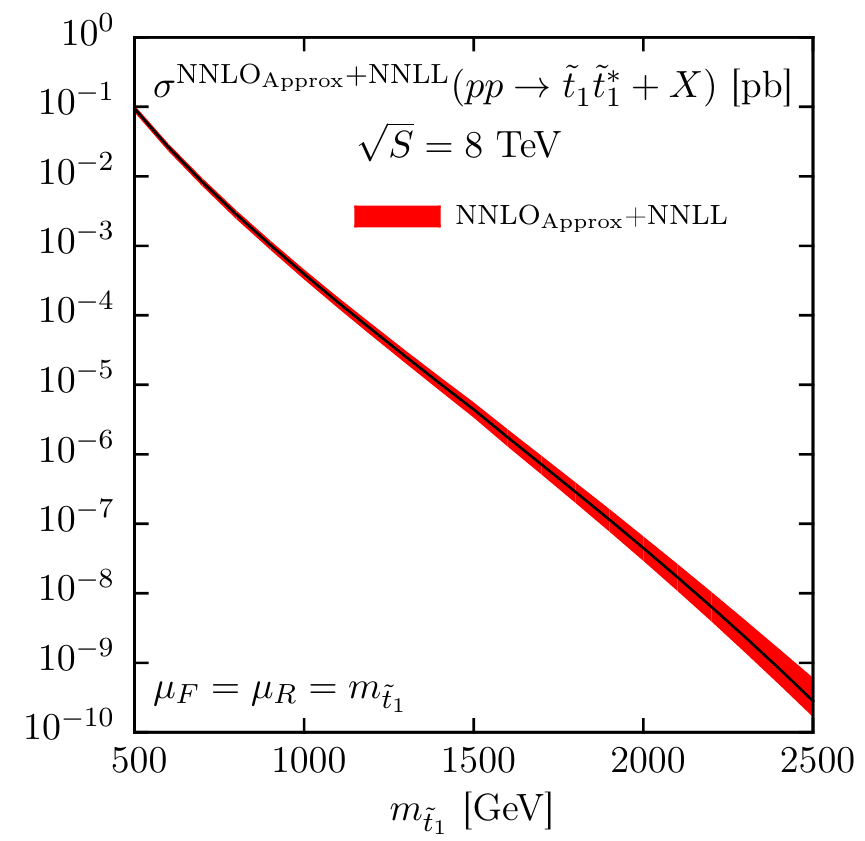}&
		\includegraphics[width=0.47\columnwidth]{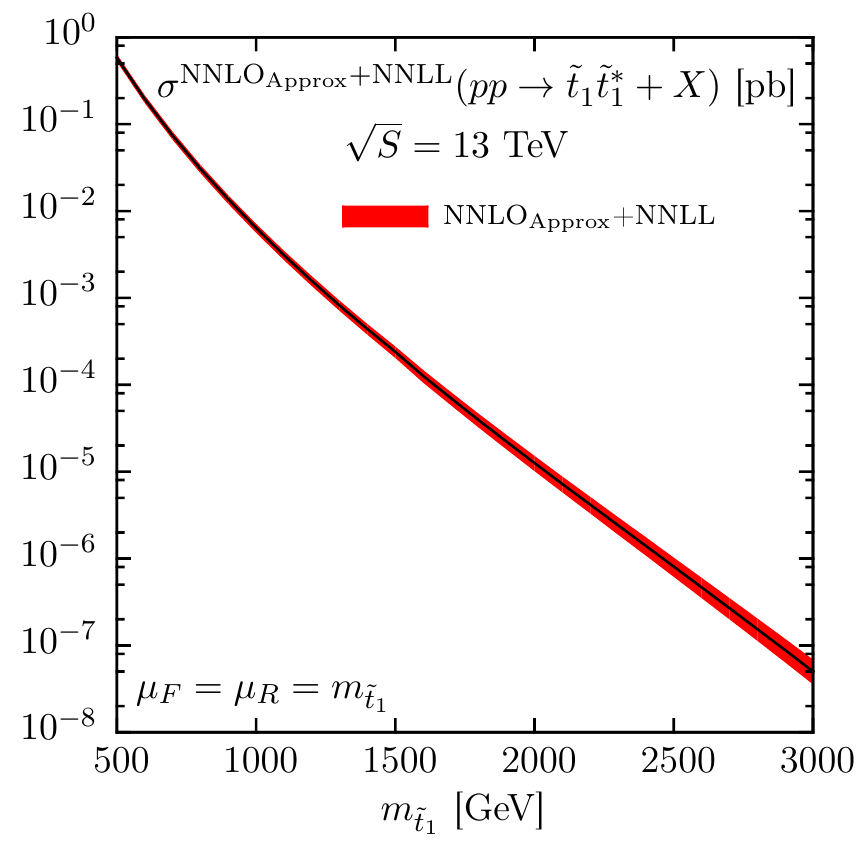}\\
		(a) & (b)
	\end{tabular}
	\caption{The NNLO$_{\mathrm{Approx}}$+NNLL cross section for stop-antistop production for LHC energies of 8 and 13 TeV, including the theoretical error band. The error band includes the 68\% C.L.\ PDF and $\alphas$ errors added quadratically, and the scale uncertainty for $m/2 \leq \mu \leq 2m$ added linearly to the combined PDF and $\alphas$ error. All parameters other than the mass of the produced stop, $m_{\sto_1}$, have been chosen according to table~\ref{tbl:benchmark}~(a).}
	\label{fig:st_tot}
\end{figure}

In figure~\ref{fig:st_tot} and table~\ref{tbl:xsections}, the total cross section for the production of $\sto_1\sto_1^*$ pairs is shown including the 68\% C.L.\ PDF and $\alphas$ uncertainties as well as the theoretical scale uncertainty for $m/2 \leq \mu \leq 2m$. While the PDF and $\alphas$ uncertainties are added in quadrature, the scale uncertainty is added linearly to the cross section. At low masses, all the three sources of uncertainty are of roughly the same size (a few per cent). The PDF and $\alphas$ uncertainties start dominating the total error above a stop mass of 1 TeV, up to individual uncertainties of 50-90\%. The scale uncertainty is roughly independent of the mass, and amounts to about 5\%. It will be analysed more in detail later in this work. A comparison to the NLO+NLL numbers from \texttt{NLL-fast} \cite{nllfast} shows that for the scale uncertainty as defined above, the effects are roughly of the same size, with a slight reduction of the uncertainty at smaller masses.

\begin{table}
	\centering
	\begin{tabular}{|c||c|c|}
		\hline
		& \multicolumn{2}{c|}{$\sigma^\mathrm{NLO+NLL}$ [pb]}\\
		$m_{\sto_1}$ [GeV] & 8 TeV & 13 TeV\\\hline\hline
		500 & $86.9^{+(8.8+6.7)\%}_{-(9.2+7.4)\%} \times 10^{-3}$ & $53.4^{+(9.2+4.6)\%}_{-(8.6+5.5)\%} \times 10^{-2}$\\
		1200 & $57.7^{+(6.1+15.1)\%}_{-(7.7+14.5)\%} \times 10^{-6}$ & $14.8^{+(6.7+9.0)\%}_{-(7.7+9.2)\%} \times 10^{-4}$\\
		2000 & $40.1^{+(1.9+30.4)\%}_{-(5.0+25.9)\%} \times 10^{-9}$ & $12.1^{+(3.6+14.8)\%}_{-(6.2+14.0)\%} \times 10^{-6}$\\\hline
	\end{tabular}
	\\[0.1cm]
	\begin{tabular}{|c||c|c|}
		\hline
		& \multicolumn{2}{c|}{$\sigma^\mathrm{NNLO_{Approx}+NNLL}$ [pb]}\\
		$m_{\sto_1}$ [GeV] & 8 TeV & 13 TeV\\\hline\hline
		500 & $94.1^{+(4.5+6.9)\%}_{-(5.1+6.6)\%} \times 10^{-3}$ & $58.1^{+(5.3+4.8)\%}_{-(4.6+5.1)\%} \times 10^{-2}$\\
		1200 & $61.8^{+(3.8+16.2)\%}_{-(5.5+13.2)\%} \times 10^{-6}$ & $15.7^{+(4.1+9.0)\%}_{-(5.0+7.9)\%} \times 10^{-4}$\\
		2000 & $44.6^{+(6.7+41.1)\%}_{-(8.7+25.9)\%} \times 10^{-9}$ & $12.6^{+(4.8+15.9)\%}_{-(6.5+12.8)\%} \times 10^{-6}$\\\hline
	\end{tabular}
	\caption{Total cross section predictions for stop-antistop production for centre-of-mass energies of $\sqrt{S} = 8$ and $\sqrt{S} =13$ TeV, shown in the format $\sigma\pm\Delta\sigma_\mu[\%]\pm\Delta\sigma_{\mathrm{PDF}+\alphas}[\%]$ with $\Delta\sigma_\mu$ being the scale uncertainty and $\Delta\sigma_{\mathrm{PDF}+\alphas}$ the combined PDF+y$\alphas$ uncertainty. For comparison, both the publicly available results for NLO+NLL as well as the new NNLO$_\mathrm{Approx}$+NNLL predictions are presented.}
	\label{tbl:xsections}
\end{table}

\begin{figure}[t]
	\centering
	\begin{tabular}{cc}
		\includegraphics[width=0.47\columnwidth]{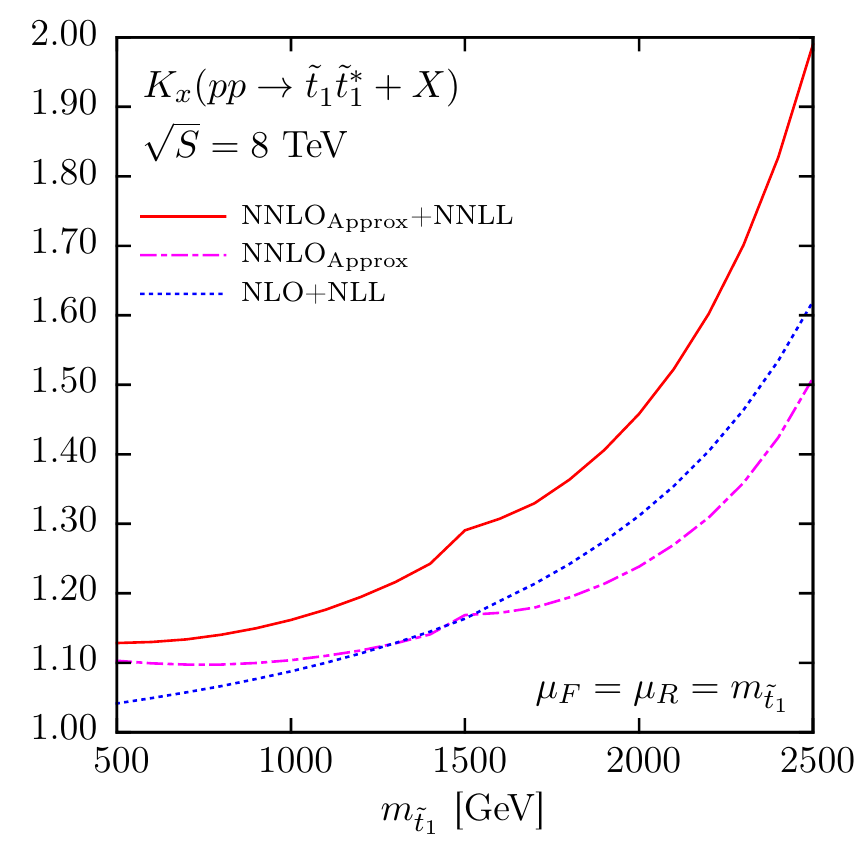}&
		\includegraphics[width=0.47\columnwidth]{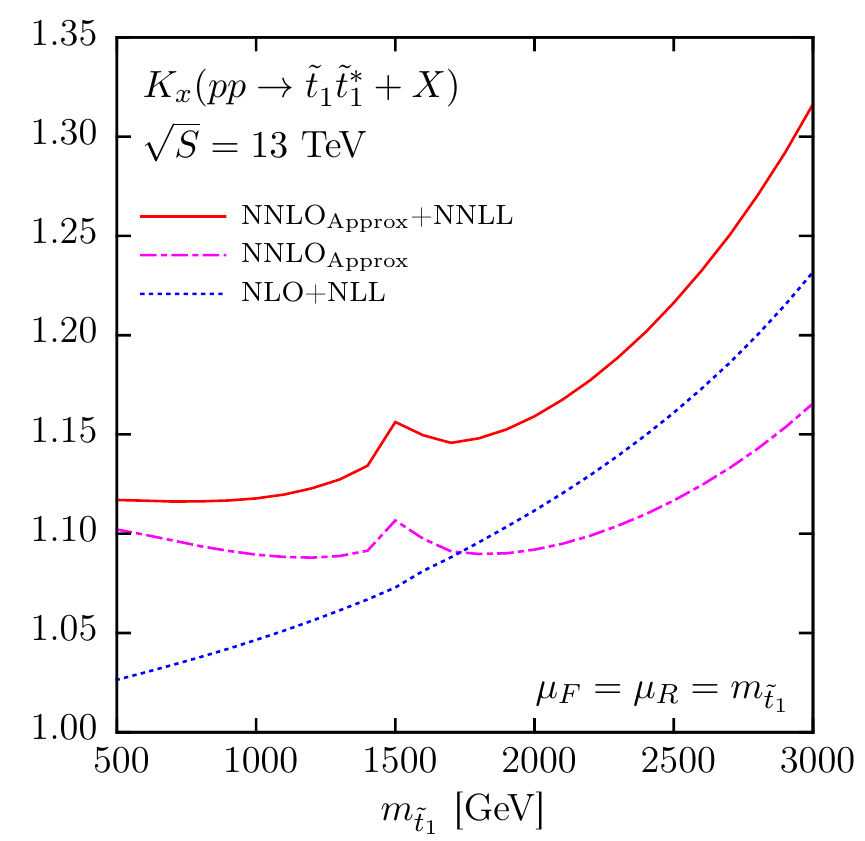}\\
		(a) & (b)
	\end{tabular}
	\caption{The $K$ factor for different orders, as explained in the text. While only the stop-1 mass is being varied, all other parameters are set to the values in table~\ref{tbl:benchmark}~(a). The renormalisation and factorisation scales have been set to the stop-1 mass.}
	\label{fig:st_k}
\end{figure}

To estimate the relative size of the resummed corrections with respect to NLO, the $K$~factor
\begin{align*}
	K_x = \frac{\sigma^x}{\sigma^{\mathrm{NLO}}}
\end{align*}
is defined and shown in figure~\ref{fig:st_k} for centre-of-mass energies of 8 and 13 TeV. The NNLL predictions are always higher than NLO, and with their ratio becoming larger as the stop mass increases. For 8 TeV, the increase is around 10\% at low stop masses and can become as large as almost 100\% at high stop masses, as expected when the production is taking place closer to the hadronic threshold, $m_{\sto_1} \approx \sqrt{S}/2$. The NNLL corrections are furthermore always larger than the previously known NLL corrections. The difference between the NNLO$_\mathrm{Approx}$ and NNLO$_\mathrm{Approx}$+NNLL predictions indicates the importance of the higher-order logarithms and their cross terms with the Coulomb and hard-matching coefficients, in particular at high stop masses close to threshold. For 13 TeV, the qualitative behaviour of the $K$ factors is similar to 8 TeV, albeit smaller at the same stop mass due to the hadronic threshold being further away. The small peak near $m_{\sto_1}\approx 1500$ GeV stems from the behaviour of the hard-matching coefficient for the $gg$ initial state, in particular due to the superposition of two effects where the virtual corrections develop a threshold for $m_{\sto_1} = m_\glu$ and $m_{\sto_1}=m_t+m_\glu$ leading to the observed behaviour close to these masses\footnote{For the hard-matching coefficient of the similar process of light-flavoured squark-antisquark production $\mathcal{C}^{(1)}_{gg\to\squ\bar{\squ},I}$ (see appendix B of \cite{Beenakker:2013mva}, p.\ 17), it has been checked that this behaviour is also arising at a similar mass, with mainly the $\log^2\left(x_{\glu\glu}(4m_\squ^2)\right)$ (with the argument $x_{\glu\glu}(4m_\squ^2) = \frac{\sqrt{1-m_\glu^2/m_\squ^2}-1}{\sqrt{1-m_\glu^2/m_\squ^2}+1}$) and the $\mathrm{Li}_2\left(\frac{m_\squ^2}{m_\glu^2}\right)$ terms stemming from the two thresholds mentioned above being responsible for it.}. The similar hard-matching coefficient of light-flavoured squark-antisquark production exhibits the same behaviour when the mass of the produced squarks is equal to the gluino mass, see \cite{Beenakker:2011sf}.

\begin{figure}[t]
	\centering
	\begin{tabular}{cc}
		\includegraphics[width=0.47\columnwidth]{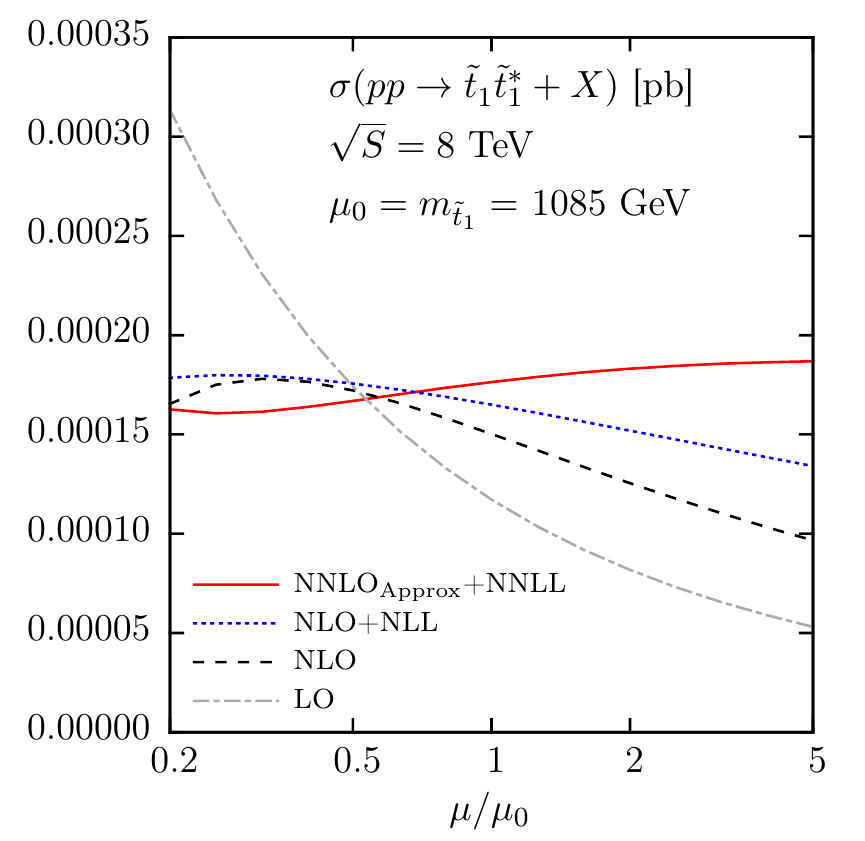}&
		\includegraphics[width=0.47\columnwidth]{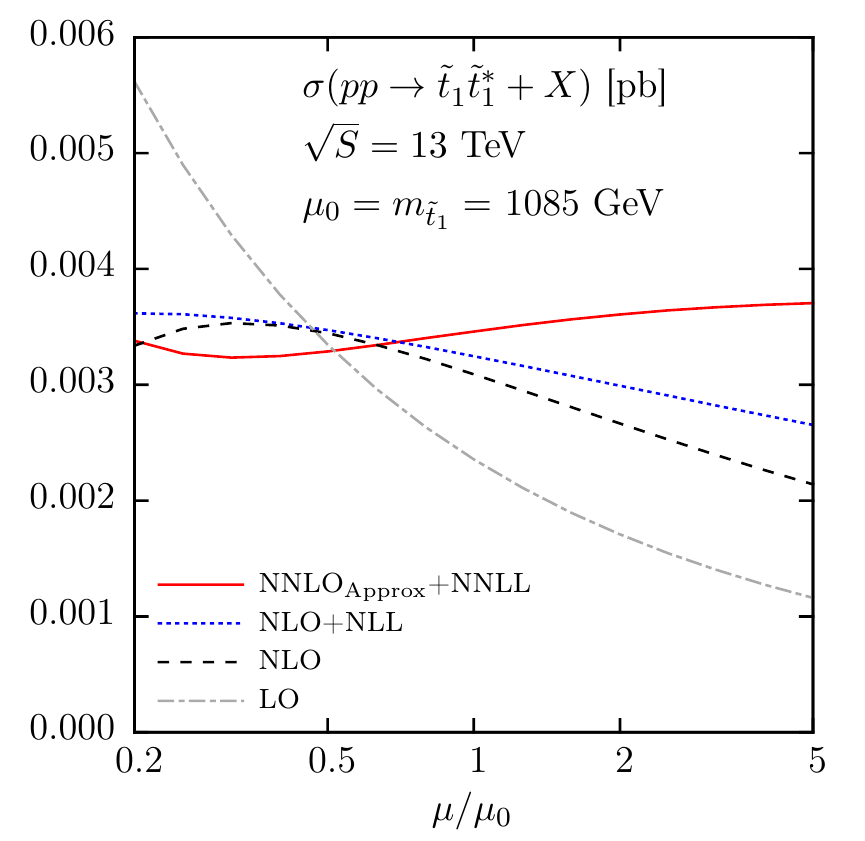}\\
		(a) & (b)
	\end{tabular}
	\caption{Scale dependence of the cross section for different orders. All masses and parameters are set to the values from table~\ref{tbl:benchmark}~(a). The central scale is chosen to be the same as the stop-1 mass.}
	\label{fig:st_mu}
\end{figure}

Scale dependence in general can give an estimate of the size of higher-order corrections that are yet to be included in the calculations. As more terms in the perturbative series are taken into account, the higher-order scale logarithms should decrease the overall dependence on the renormalisation and factorisation scales. In figure~\ref{fig:st_mu}, the scale dependence for several accuracies of fixed-order and resummed cross sections is shown, again for the two centre-of-mass energies of 8 and 13 TeV. The scale is varied around the central value of $\mu_0 = m_{\sto_1} = 1085$ GeV from $\mu = \mu_0/5$ to $\mu = 5\mu_0$. While there is already a notable decrease of the scale dependence at the NLL level with respect to NLO, the behaviour furthermore improves due to the NNLL contributions. It should be noted that even though the scale uncertainty numbers from table~\ref{tbl:xsections} suggest that the scale dependence increases for high stop masses when adding the NNLL corrections, figure~\ref{fig:st_mu} reveals that the overall slope of the curve becomes flatter than NLO+NLL, in particular towards higher scales. Additionally, the scale uncertainty computed at the two values $\mu = \mu_0/2$ and $\mu = 2\mu_0$ can become very small for NLO or NLO+NLL when the mass of the stops leads to the turn-over point of the scale dependence (as can be seen in figure~\ref{fig:st_mu} for NLO in between $\mu = 0.2\mu_0$ and $\mu = 0.5\mu_0$) being close to the central scale $\mu = \mu_0$.

In comparison to the previous results at NLO+NLL, the NNLO$_\mathrm{Approx}$+NNLL prediction now shows a slight upward slope. In section~\ref{sub:sb_comp}, this behaviour of the scale dependence is compared to the similar production of light-flavoured squark-antisquark pairs as studied in \cite{Beenakker:2011sf,Beenakker:2014sma}, and it will be seen that the difference in the behaviour mainly stems from the absence of the singlet colour channel for the $q\bar{q}$ initial state.

\begin{table}
	\centering
	\begin{tabular}{|c||c|c|}
		\hline
		Parameter & Value & $\sigma \times 10^{-4}$ [pb]\\\hline\hline
		40.2.5 & -- & 34.60\\\hline
		\multirow{4}{*}{$m_{\sto_2}$} & 100 GeV & 34.65\\
		& 500 GeV & 34.60\\
		& 2000 GeV & 34.60\\
		& 5000 GeV & 34.60\\\hline
		\multirow{4}{*}{$m_{\glu}$} & 100 GeV & 33.15\\
		& 500 GeV & 31.98\\
		& 2000 GeV & 34.63\\
		& 5000 GeV & 34.80\\\hline
		\multirow{4}{*}{$m_{\squ}$} & 100 GeV & 34.85\\
		& 500 GeV & 34.46\\
		& 2000 GeV & 34.76\\
		& 5000 GeV & 35.02\\\hline
		\multirow{3}{*}{$\sin 2\theta_\sto$} & 1 & 34.60\\
		& 0 & 34.67\\
		& -1 & 34.95\\\hline
	\end{tabular}
	\caption{Parameter dependence of stop-antistop production at NNLO$_\mathrm{Approx}$+NNLL for $\sqrt{S} = 13$ TeV. All parameters are chosen corresponding to benchmark point 40.2.5, and then varied individually for the given values.}
	\label{tbl:param_dep}
\end{table}
\begin{figure}[t]
	\centering
	\includegraphics[width=0.47\columnwidth]{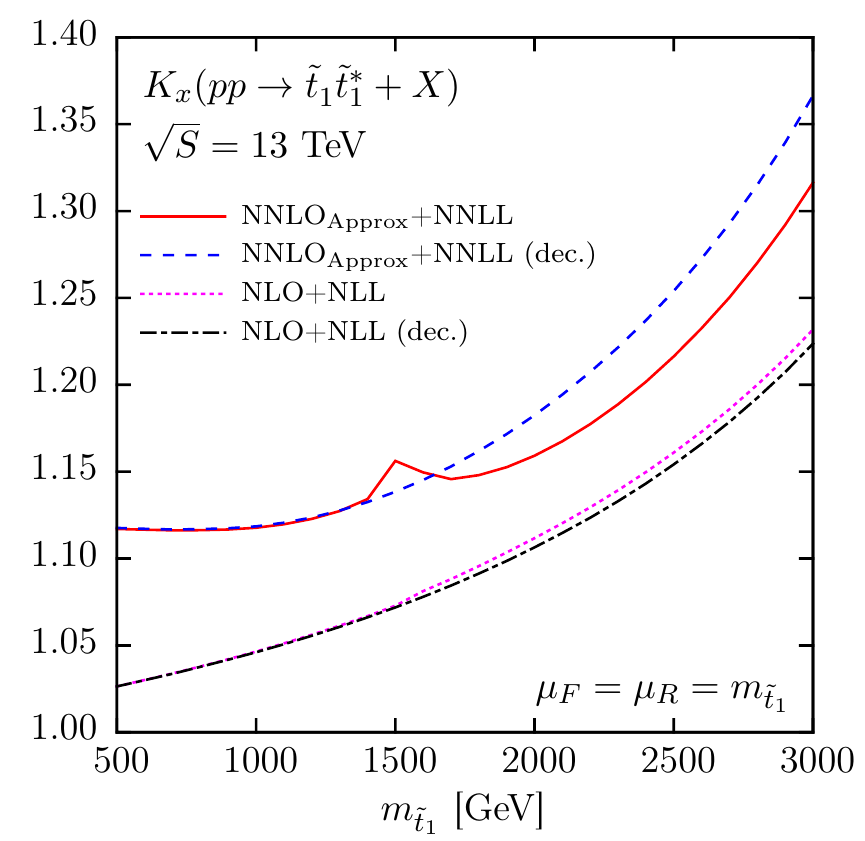}
	\caption{Comparison of the $K$ factor when choosing the additional stop parameters according to table~\ref{tbl:benchmark}~(a) (solid red for NNLO$_\mathrm{Approx}$+NNLL, pink dotted for NLO+NLL) and when decoupling all masses except for $m_{\sto_1}$ (dashed blue for NNLO$_\mathrm{Approx}$+NNLL, black dash-dotted for NLO+NLL). For the lines corresponding to the decoupling limit, also the NLO cross section appearing in the denominator of the $K$ factor is computed with the decoupled parameters.}
	\label{fig:st_dec}
\end{figure}

The dependence of the NNLO$_\mathrm{Approx}$+NNLL results on the additional SUSY parameters appearing first at NLO is shown in table~\ref{tbl:param_dep}. The different parameters are first set to the default values of table~\ref{tbl:benchmark}~(a), and subsequently varied separately. While the variation of the heavier $\sto_2$ mass $m_{\sto_2}$ hardly influences the cross section (even when the $\sto_2$ is actually lighter than the $\sto_1$), a change of the mixing angle or the degenerate light-flavoured squark mass leads to a deviation of at most 1\% from using the default parameter values. The main impact comes from the variation of the gluino mass which changes the cross section by several per cent. In figure~\ref{fig:st_dec}, one exemplary choice of parameters where all other squarks and gluinos except for the produced light stop are decoupled\footnote{i.e.\ their masses are considered much higher than $m_{\sto_1}$. For concreteness, the decoupled parameters in the plot are $m_{\squ} = 10$~TeV, $m_{\sto_2} = 10.01$~TeV, $m_{\glu} = 10.02$~TeV.} is shown as a comparison to the $K$ factor for the default parameter values. At high $m_{\sto_1}$, the $K$ factor for NNLO$_\mathrm{Approx}$+NNLL differs by as much as 5\%, accentuating the observation that the choice of a particular SUSY model and benchmark point may have a noticeable effect on the stop production cross sections. In the comparison to the NLO+NLL lines, it can be seen that this large difference at high masses stems from the hard-matching coefficient as the only ingredient at NNLL accuracy that depends on the additional parameters, in particular the gluino mass. More precisely, in the NLO+NLL $K$ factor the decoupling effects enter through $\sigma_{\rm NLO}$ in the denominator, therefore being of ${\cal O}(\alpha_s^3)$ and having the reversed sign of $\sigma_{\rm NLO}$. The NNLO$_{\rm Approx}$ +NNLL results on the other hand receive additional ${\cal O}(\alpha_s^2)$ contributions from one-loop Coulomb times one-loop hard matching coefficients, resulting in larger decoupling effects with the same sign as $\sigma_{\rm NLO}$.

\subsection{Comparison to squark-antisquark production}\label{sub:sb_comp}
\begin{figure}[t]
	\centering
	\begin{tabular}{cc}
		\includegraphics[width=0.47\columnwidth]{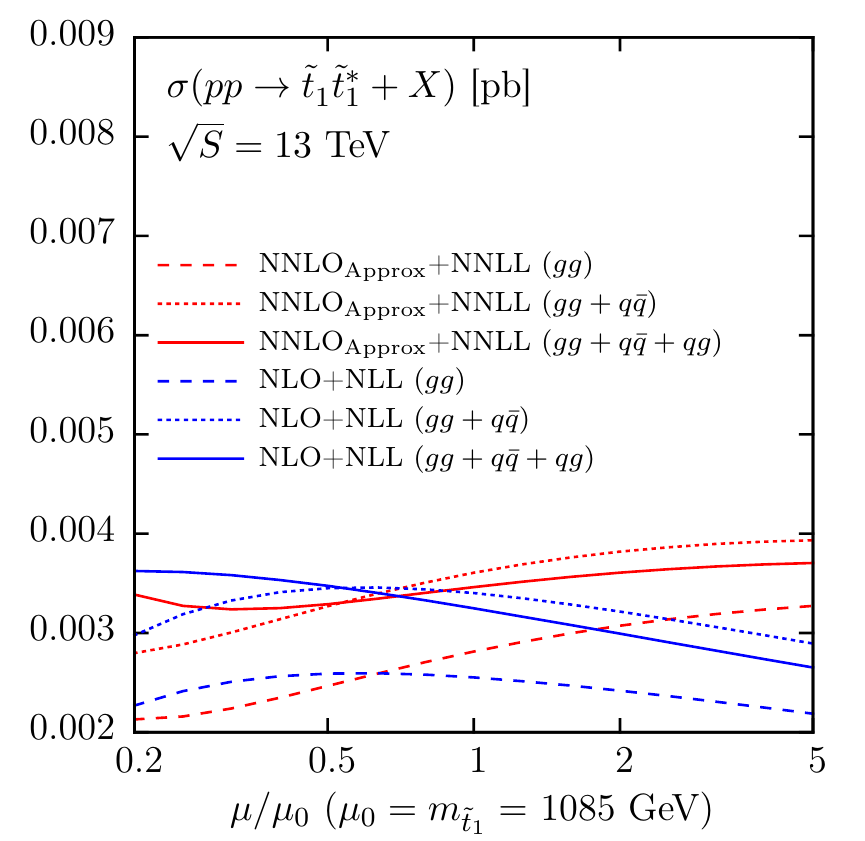}&
		\includegraphics[width=0.47\columnwidth]{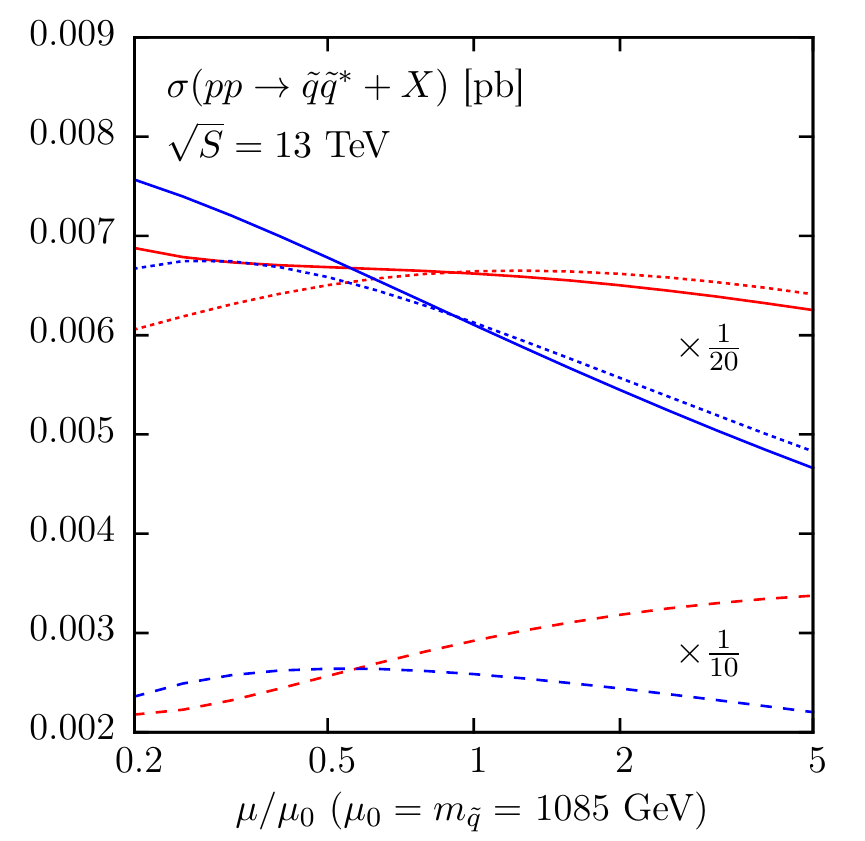}\\
		(a) & (b)\\
	\end{tabular}
	\includegraphics[width=0.47\columnwidth]{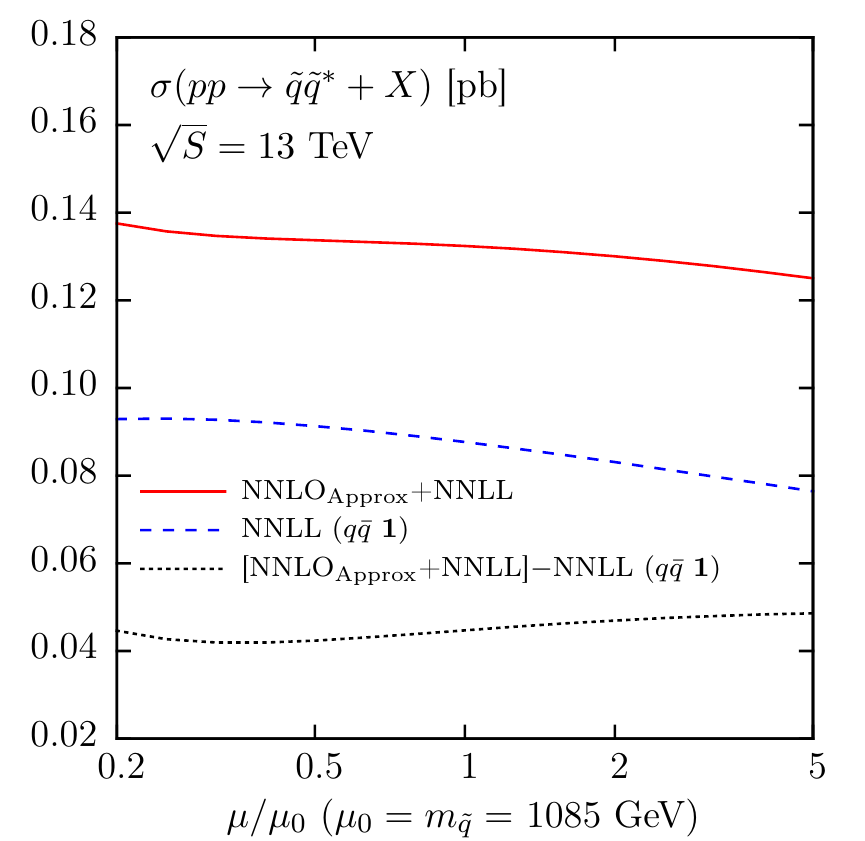}\\
	(c)
	\caption{Scale dependence for stop-antistop (a) and light-flavoured squark-antisquark production (b) at 13 TeV, split up into the initial-state contributions, both for NLO+NLL as well as NNLO$_{\mathrm{Approx}}$+NNLL. The labelling of the lines in (b) is the same as in (a). It should be noted that for an easier comparison to (a), the lines of (b) have been divided by a factor of 10 for the $gg$ initial state (dashed lines), and by a factor of 20 for the $gg+q\bar{q}$ (dotted lines) and $gg+q\bar{q}+qg$ initial states (solid lines), as indicated on the plot. In (c), the contribution from the singlet channel of the $q\bar{q}$ initial state to $\squ\squ^*$ is shown, see the accompanying text for further explanations. The central scales as well as the masses of the produced particles are set in both cases to the $\sto_1$ mass of table~\ref{tbl:benchmark}~(a) for easier comparison.}
	\label{fig:st_sb_comp}
\end{figure}

Figure~\ref{fig:st_sb_comp} shows the scale dependence of stop-antistop and light-flavoured squark-antisquark production at 13 TeV, split up into the different initial-state channels $gg$, $q\bar{q}$, and $qg$, where the latter only appears starting from NLO and does not receive any contributions from resummation. It is interesting to compare the two processes, as they are very similar in their production mechanisms, yet exhibit a slightly different behaviour when it comes to their scale dependence.

In figure~\ref{fig:st_sb_comp} (a), the scale dependence of $\sto\sto^*$ is plotted, both for the NLO+NLL as well as the NNLO$_{\mathrm{Approx}}$+NNLL accuracies. It can be seen that the contributions from the $gg$ initial state differ at these two levels: for NNLO$_{\mathrm{Approx}}$+NNLL, $gg$ shows a clear upward slope. Adding the threshold-suppressed $q\bar{q}$ contributions does not influence the behaviour significantly, and only increases the cross section in both cases. It is clear that the $gg$ and $q\bar{q}$ contributions actually do not differ that much, with the former being larger by a modest factor at the given value for the stop mass. The $qg$ contribution is responsible for an increase of the cross section at lower scales, and a decrease at higher scales. The solid red line shows the new NNLO$_\mathrm{Approx}$+NNLL prediction as presented in the previous section. At NNLO$_{\mathrm{Approx}}$+NNLL, the upward slope therefore originates from the behaviour of the dominant $gg$ channel.

Now, figure~\ref{fig:st_sb_comp} (b) is the analogous plot for $\squ\squ^*$. While the $gg$ contributions behave in exactly the same way (as they are, in fact, the same aside from the factor $2n_f$), the most notable difference is the size of the $q\bar{q}$ contributions, which are clearly dominant in this case. The $qg$ contribution again mainly leads to an increase of the cross section at low scales. Going a bit more into detail about the comparison of the $q\bar{q}$ initial state for both $\sto\sto^*$ and $\squ\squ^*$, the difference at LO in the first case is the absence of the $t$-channel graph with gluino exchange. This graph is solely responsible for the singlet contribution, and also to a lesser degree for additional terms to the octet (compare equations \eqref{qqbstst1} and \eqref{qqbstst8} to e.g.\ equations (9) and (10) of \cite{Kulesza:2009kq}). Even though the remaining octet term for $\sto\sto^*$ is in a $p$-wave channel, it is still comparable in size with the $gg$ contributions. The singlet channel, missing in stop production and present in the light-flavoured squark-antisquark predictions, is clearly dominant; figure~\ref{fig:st_sb_comp} (b) shows that it is the cause of the large increase of the latter over the former. Furthermore, the contributions from the $q\bar{q}$ initial state lead to the downward slope of the scale dependence at the NLL level. The additional NNLL corrections lead to an almost flat scale dependence with a slight downward slope towards higher scales remaining.

Figure~\ref{fig:st_sb_comp} (c) shows the NNLO$_{\mathrm{Approx}}$+NNLL prediction for $\squ\squ^*$ (red solid line), the purely resummed contributions for the singlet channel of the $q\bar{q}$ initial state labelled ``NNLL ($q\bar{q}~\mathbf{1}$)'' (blue dashed line), and the latter subtracted from the former (black dotted line). This last line is an approximation for removing all singlet-$q\bar{q}$ contributions from the $\squ\squ^*$~cross section\footnote{It is an approximation, as it is not possible to switch off only certain colour channels in \texttt{Prospino}, so only the threshold-enhanced NLO terms of the singlet-$q\bar{q}$, contained within NNLL ($q\bar{q}~\mathbf{1}$), are removed.}. It is clear that doing this first of all leads to a notable decrease in the cross section, and secondly changes the scale behaviour to exhibit an upward slope, as observed for $\sto\sto^*$. In conclusion, it can therefore be said that the absence of the singlet $q\bar{q}$-channel leads to a dominance of the $gg$ initial state, resulting in an upward slope of the scale dependence of the stop-antistop cross section at the NNLO$_{\mathrm{Approx}}$+NNLL accuracy\footnote{It should be noted that this behavior of the scale dependence is similar to the one observed for gluino-pair production (see \cite{Beenakker:2014sma}), dominated by gluon-channel contributions.}.

\subsection{Comparison to SCET results}
A study of stop-antistop production including threshold resummation at NNLL accuracy has been done in \cite{Broggio:2013cia} in the framework of SCET. Two different ways of approaching a threshold with different threshold definitions were used, commonly referred to as the so-called ``pair invariant mass'' and the ``one particle inclusive'' kinematics. It is interesting to see how the results of \cite{Broggio:2013cia} compare to the one presented in this paper.

\begin{table}
	\centering
	\begin{tabular}{|c||c|c|}
		\hline
		& \multicolumn{2}{c|}{8 TeV}\\
		$m_{\sto_1}$ [GeV] & 500 & 1087.17 \\\hline\hline
		$(\sigma\pm\Delta\sigma_\mu)_\mathrm{NLO}$ [pb] & $83.4^{+10.5}_{-12.1} \times 10^{-3}$ & $14.7^{+2.1}_{-2.4} \times 10^{-5}$\\\hline
		$(\sigma\pm\Delta\sigma_\mu)_\mathrm{NLO+NNLL}$ [pb] from \cite{Broggio:2013cia} & $79.9^{+6.2}_{-3.9} \times 10^{-3}$ & $15.2^{+1.7}_{-0.9} \times 10^{-5}$\\\hline
		$(\sigma\pm\Delta\sigma_\mu)_{\mathrm{NLO+NNLL}_{h+\mathrm{Coul}(1)}}$ [pb] & $85.2^{+0.8}_{-0.2} \times 10^{-3}$ & $15.9^{+0.2}_{-0.2} \times 10^{-5}$\\\hline
		$(\sigma\pm\Delta\sigma_\mu)_{\mathrm{NLO+NNLL}}$ [pb] & $86.4^{+0.2}_{-0.0} \times 10^{-3}$ & $16.2^{+0.3}_{-0.3} \times 10^{-5}$\\\hline
		$(\sigma\pm\Delta\sigma_\mu)_{\mathrm{NNLO_{Approx}+NNLL}}$ [pb] & $94.1^{+4.3}_{-4.8} \times 10^{-3}$ & $17.3^{+0.7}_{-0.9} \times 10^{-5}$\\\hline
	\end{tabular}
	\\[0.1cm]
	\begin{tabular}{|c||c|c|}
		\hline
		& \multicolumn{2}{c|}{14 TeV}\\
		$m_{\sto_1}$ [GeV] & 500 & 1087.17 \\\hline\hline
		$(\sigma\pm\Delta\sigma_\mu)_\mathrm{NLO}$ [pb] & $66.5^{+7.5}_{-8.5} \times 10^{-2}$ & $44.2^{+4.9}_{-5.9} \times 10^{-4}$\\\hline
		$(\sigma\pm\Delta\sigma_\mu)_\mathrm{NLO+NNLL}$ [pb] from \cite{Broggio:2013cia} & $62.9^{+3.5}_{-3.2} \times 10^{-2}$ & $43.1^{+3.5}_{-1.8} \times 10^{-4}$\\\hline
		$(\sigma\pm\Delta\sigma_\mu)_{\mathrm{NLO+NNLL}_{h+\mathrm{Coul}(1)}}$ [pb] & $67.2^{+1.5}_{-0.4} \times 10^{-2}$ & $45.3^{+0.1}_{-0.0} \times 10^{-4}$\\\hline
		$(\sigma\pm\Delta\sigma_\mu)_{\mathrm{NLO+NNLL}}$ [pb] & $68.0^{+1.0}_{-0.1} \times 10^{-2}$ & $45.9^{+0.3}_{-0.2} \times 10^{-4}$\\\hline
		$(\sigma\pm\Delta\sigma_\mu)_{\mathrm{NNLO_{Approx}+NNLL}}$ [pb] & $74.3^{+4.0}_{-3.4} \times 10^{-2}$ & $49.3^{+2.2}_{-2.4} \times 10^{-4}$\\\hline
	\end{tabular}
	\caption{Stop-antistop production cross sections for different accuracies at centre-of-mass energies of $\sqrt{S} = 8, 14$ TeV using MSTW2008 PDF sets. Only the scale variation uncertainties are shown. Refer to the accompanying text for more explanation on these uncertainties. For all numbers, the parameter values from table~\ref{tbl:benchmark}~(b) are used.}
	\label{tbl:xsections_scet}
\end{table}

\begin{figure}[t]
	\centering
	\includegraphics[width=0.47\columnwidth]{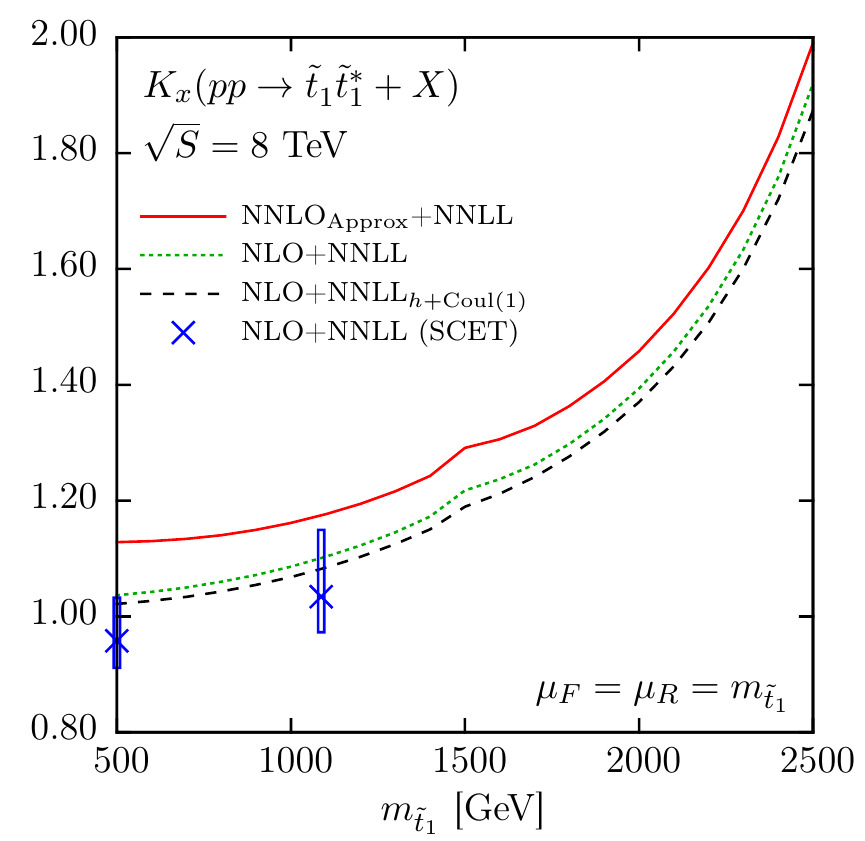}
	\caption{Comparison of the $K$ factor to the SCET results \cite{Broggio:2013cia}. Shown is the NNLO$_{\mathrm{Approx}}$+NNLL prediction of this paper (solid red), the NLO+NNLL$_{h+\mathrm{Coul}(1)}$ cross section mimicking the method of \cite{Broggio:2013cia} (dashed black), the NLO+NNLL cross section including $\mathcal{O}(\alphas^2)$ terms in the matching coefficient (dotted green), and two points for the SCET results including their uncertainties from table~\ref{tbl:xsections_scet}. All lines and points in this plot have been calculated using the values of table~\ref{tbl:benchmark}~(b).}
	\label{fig:st_k_scet}
\end{figure}

Table~\ref{tbl:xsections_scet} shows the total cross sections for various accuracies and matchings using the parameter points of table~\ref{tbl:benchmark}~(b), also plotted in figure~\ref{fig:st_k_scet} for a wider stop mass range. The numbers are compared for $\sqrt{S} = 8$ TeV, using the MSTW2008 PDF sets. Only the theoretical scale uncertainties are shown. It should be noted that the authors of \cite{Broggio:2013cia} obtain their errors from varying all the scales individually for the two different kinematics and then computing the envelope of the minimum and maximum values, leading to significantly larger uncertainties than for the numbers from this paper. Nonetheless, it is clear that the results of \cite{Broggio:2013cia} agree well with the ones presented here within the given uncertainties, taking into account the different matching procedure and disabling $\mathcal{O}(\alphas^2)$ terms in the matching coefficients $C_{ij\to\sto\sto^*,I}$. This is important as it shows that with different methods of defining the threshold region, similar results can be obtained. It should be noted, however, that for a stop mass of 500 GeV, the NNLL corrections of the SCET formalism decrease the cross sections with respect to NLO, albeit being compatible with a $K$ factor larger than 1 within uncertainties. The same behaviour is not seen for the results presented in this paper. The main difference between the NNLO$_{\mathrm{Approx}}$+NNLL predictions of this work and the results of \cite{Broggio:2013cia} seems to originate from different accuracies employed in the matching procedure, as discussed at the beginning of section~\ref{s:numerics}.

%%%%%%%%%%%%%%%%%%%%%%%%%%%%%%%%%%%%%%%%%%%
\section{Conclusions and outlook}\label{s:conclusion}
%%%%%%%%%%%%%%%%%%%%%%%%%%%%%%%%%%%%%%%%%%%
This work presents results for supersymmetric stop-antistop production at the LHC with $\sqrt{S} =$ 8 and 13 TeV including corrections due to soft-gluon resummation at NNLL matched to an approximation of NNLO. The calculations have been done in the Mellin-space approach, and update the previous NLO+NLL results of \cite{Beenakker:2010nq}. The NNLO$_\mathrm{Approx}$+NNLL cross section is significantly enhanced with respect to the NLO+NLL and also the fixed-order NNLO$_\mathrm{Approx}$ results. This emphasizes the importance of the higher-order logarithms, the higher-order corrections to the hard function, and the Coulomb contributions at high stop masses close to the hadronic threshold. In particular the latter contributions strongly influence the behaviour of the cross section close to threshold, as was seen in the analysis of the NLO scaling function. For 8 TeV and a stop mass of 2.5 TeV, the NNLO$_\mathrm{Approx}$+NNLL results are increased by roughly 40\% compared to NLO+NLL, whereas the increase for 13 TeV and a stop mass of 3 TeV is only 10\% due to the latter case being further away from the hadronic threshold. It was also seen that due to the inclusion of the hard-matching coefficients, the NNLL results now depend more strongly than the NLL results on the specific SUSY parameters that appear as a consequence of the stop mixing. In particular, the hierarchy of stop and gluino masses has been found to matter.

The scale dependence of stop-antistop production is strongly influenced by the new contributions taken into account. In particular, it was seen that the scale behaviour now exhibits a slight upward slope towards higher scales, while overall still showing a decrease in scale dependence compared to the NLO+NLL results. In a detailed comparison to the similar process of light-flavoured squark-antisquark production, the absence of the singlet colour channel of the $q\bar{q}$ initial state that dominates the light-flavoured process has been found responsible for the observed scale behaviour for stop-antistop production. The scale dependence is therefore driven by the $gg$ initial state which leads to a lower value of the cross section for low scales, and a higher value for high scales.

With the availability of NNLL results for stop-antistop production in \cite{Broggio:2013cia}, it is interesting to see how the method presented in this paper compares to the results calculated in the SCET formalism. The two different kinematics of \cite{Broggio:2013cia} in which the threshold region is defined lead to predictions that are comparable within the theoretical uncertainties with the prediction presented in this work if the same classes of terms are considered. A full comparison of the dependence on the scales will be difficult due to the variety of scales in the SCET formalism which is not available in the Mellin-space method. Nonetheless, the good agreement of the available predictions, obtained in two independent approaches, confirms that with different methods of threshold resummation, similar results can be achieved.

In the near future, the results for stop production at NNLL will be published together with the other squark and gluino production processes as a public code akin to \texttt{NLL-fast} \cite{nllfast} for a centre-of-mass energy of 13 TeV to be used in experimental studies for Run II of the LHC.

%%%%%%%%%%%%%%%%%%%%%%%%%%%%%%%%%%%%%%%%%%%
\section*{Acknowledgments}
%%%%%%%%%%%%%%%%%%%%%%%%%%%%%%%%%%%%%%%%%%%
This work has been supported in part by the Netherlands Foundation for Fundamental Research of Matter (FOM) programme 156, entitled ``Higgs as Probe and Portal'', and the National Organization for Scientific Research (NWO), the BMBF Verbundprojekt 05H2015 (BMBF-FSP 104), DFG grant KU 3103/1, and the DFG research unit 2239 ``New physics at the LHC''.

\bibliographystyle{JHEP}

\begin{thebibliography}{10}

\bibitem{Wess:1973kz}
	J.~Wess and B.~Zumino,
	{\it A Lagrangian Model Invariant Under Supergauge Transformations},
	Phys.\ Lett.\ B {\bf 49} (1974) 52.

\bibitem{Wess:1974tw}
	J.~Wess and B.~Zumino,
	{\it Supergauge Transformations in Four-Dimensions},
	Nucl.\ Phys.\ B {\bf 70} (1974) 39.

\bibitem{Fayet:1976et}
	P.~Fayet,
	{\it Supersymmetry and Weak, Electromagnetic and Strong Interactions},
	Phys.\ Lett.\ B {\bf 64} (1976) 159.

\bibitem{Farrar:1978xj}
	G.~R.~Farrar and P.~Fayet,
	{\it Phenomenology of the Production, Decay, and Detection of New Hadronic States Associated with Supersymmetry},
	Phys.\ Lett.\ B {\bf 76} (1978) 575.

\bibitem{Sohnius:1985qm}
	M.~F.~Sohnius,
	{\it Introducing Supersymmetry},
	Phys.\ Rept.\  {\bf 128} (1985) 39.

\bibitem{Martin:1997ns}
	S.~P.~Martin,
	{\it A Supersymmetry primer},
	Adv.\ Ser.\ Direct.\ High Energy Phys.\  {\bf 21} (2010) 1
	[Adv.\ Ser.\ Direct.\ High Energy Phys.\  {\bf 18} (1998) 1]
	[\href{http://arxiv.org/abs/hep-ph/9709356}{\tt hep-ph/9709356}].

\bibitem{Aad:2015iea}
	G.~Aad {\it et al.} [ATLAS Collaboration],
	{\it Summary of the searches for squarks and gluinos using $ \sqrt{s}=8 $ TeV pp collisions with the ATLAS experiment at the LHC},
	JHEP {\bf 1510} (2015) 054
	[\href{http://arxiv.org/abs/1507.05525}{\tt arXiv:1507.05525}].

\bibitem{Aad:2015baa}
	G.~Aad {\it et al.} [ATLAS Collaboration],
	{\it Summary of the ATLAS experiment's sensitivity to supersymmetry after LHC Run 1 - interpreted in the phenomenological MSSM},
	[\href{http://arxiv.org/abs/1508.06608}{\tt arXiv:1508.06608}].

\bibitem{Khachatryan:2015lwa}
	V.~Khachatryan {\it et al.} [CMS Collaboration],
	{\it Search for Physics Beyond the Standard Model in Events with Two Leptons, Jets, and Missing Transverse Momentum in pp Collisions at sqrt(s) = 8 TeV},
	JHEP {\bf 1504} (2015) 124
	[\href{http://arxiv.org/abs/1502.06031}{\tt arXiv:1502.06031}].

\bibitem{Ellis:1983ed}
	J.~R.~Ellis and S.~Rudaz,
	{\it Search for Supersymmetry in Toponium Decays},
	Phys.\ Lett.\ B {\bf 128} (1983) 248.

\bibitem{Aad:2015pfx}
	G.~Aad {\it et al.} [ATLAS Collaboration],
	{\it ATLAS Run 1 searches for direct pair production of third-generation squarks at the Large Hadron Collider},
	[\href{http://arxiv.org/abs/1506.08616}{\tt arXiv:1506.08616}].

\bibitem{Chatrchyan:2013xna}
	S.~Chatrchyan {\it et al.} [CMS Collaboration],
	{\it Search for top-squark pair production in the single-lepton final state in pp collisions at $\sqrt{s}$ = 8 TeV},
	Eur.\ Phys.\ J.\ C {\bf 73} (2013) 12,  2677
	[\href{http://arxiv.org/abs/1308.1586}{\tt arXiv:1308.1586}].

\bibitem{Chatrchyan:2013mya}
	S.~Chatrchyan {\it et al.} [CMS Collaboration],
	{\it Search for top squark and higgsino production using diphoton Higgs boson decays},
	Phys.\ Rev.\ Lett.\  {\bf 112} (2014) 161802
	[\href{http://arxiv.org/abs/1312.3310}{\tt arXiv:1312.3310}].

\bibitem{Khachatryan:2015wza}
	V.~Khachatryan {\it et al.} [CMS Collaboration],
	{\it Searches for third-generation squark production in fully hadronic final states in proton-proton collisions at $ \sqrt{s} = 8$ TeV},
	JHEP {\bf 1506} (2015) 116
	[\href{http://arxiv.org/abs/1503.08037}{\tt arXiv:1503.08037}].

\bibitem{Nilles:1983ge}
	H.~P.~Nilles,
	{\it Supersymmetry, Supergravity and Particle Physics},
	Phys.\ Rept.\  {\bf 110} (1984) 1.

\bibitem{Haber:1984rc}
	H.~E.~Haber and G.~L.~Kane,
	{\it The Search for Supersymmetry: Probing Physics Beyond the Standard Model},
	Phys.\ Rept.\  {\bf 117} (1985) 75.

\bibitem{Beenakker:1997ut}
	W.~Beenakker, M.~Krämer, T.~Plehn, M.~Spira, and P.~M. Zerwas,
	{\it {Stop production at hadron colliders}},
	{\em Nucl. Phys.} {\bf B515} (1998) 3--14,
	[\href{http://arxiv.org/abs/hep-ph/9710451}{\tt hep-ph/9710451}].

\bibitem{Hollik:2007wf}
	W.~Hollik, M.~Kollar and M.~K.~Trenkel,
	{\it Hadronic production of top-squark pairs with electroweak NLO contributions},
	JHEP {\bf 0802} (2008) 018
	[\href{http://arxiv.org/abs/0712.0287}{\tt arXiv:0712.0287}].

\bibitem{Germer:2014jpa}
	J.~Germer, W.~Hollik, J.~M.~Lindert and E.~Mirabella,
	{\it Top-squark pair production at the LHC: a complete analysis at next-to-leading order,}
	JHEP {\bf 1409} (2014) 022
	[\href{http://arxiv.org/abs/1404.5572}{\tt arXiv:1404.5572}].

\bibitem{Sterman:1986aj}
	G.~F. Sterman,
	{\it {Summation of Large Corrections to Short Distance Hadronic Cross-Sections}},
	{\em Nucl. Phys.} {\bf B281} (1987) 310.

\bibitem{Catani:1989ne}
	S.~Catani and L.~Trentadue,
	{\it {Resummation of the QCD Perturbative Series for Hard Processes}},
	{\em Nucl. Phys.} {\bf B327} (1989) 323.

\bibitem{Bonciani:1998vc}
	R.~Bonciani, S.~Catani, M.~L. Mangano, and P.~Nason, {\it {NLL Resummation of the Heavy-Quark Hadroproduction Cross-Section}},
	{\em Nucl. Phys.} {\bf B529} (1998) 424
	[\href{http://arxiv.org/abs/hep-ph/9801375}{\tt hep-ph/9801375}].

\bibitem{Contopanagos:1996nh}
	H.~Contopanagos, E.~Laenen, and G.~Sterman, {\it {Sudakov Factorization and Resummation}},
	{\em Nucl. Phys.} {\bf B484} (1997) 303
	[\href{http://arxiv.org/abs/hep-ph/9604313}{\tt hep-ph/9604313}].

\bibitem{Kidonakis:1998bk}
	N.~Kidonakis, G.~Oderda, and G.~Sterman, {\it {Threshold Resummation for Dijet Cross Sections}},
	{\em Nucl. Phys.} {\bf B525} (1998) 299
	[\href{http://arxiv.org/abs/hep-ph/9801268}{\tt hep-ph/9801268}].

\bibitem{Kidonakis:1998nf}
	N.~Kidonakis, G.~Oderda, and G.~Sterman, {\it {Evolution of Color Exchange in {QCD} Hard Scattering}},
	{\em Nucl. Phys.} {\bf B531} (1998) 365
	[\href{http://arxiv.org/abs/hep-ph/9803241}{\tt hep-ph/9803241}].

\bibitem{Kulesza:2008jb}
	A.~Kulesza and L.~Motyka,
	{\it Threshold resummation for squark-antisquark and gluino-pair production at the LHC},
	Phys.\ Rev.\ Lett.\  {\bf 102} (2009) 111802
	[\href{http://arxiv.org/abs/0807.2405}{\tt arXiv:0807.2405}].

\bibitem{Kulesza:2009kq}
	A.~Kulesza and L.~Motyka,
	{\it Soft gluon resummation for the production of gluino-gluino and squark-antisquark pairs at the LHC},
	Phys.\ Rev.\ D {\bf 80} (2009) 095004
	[\href{http://arxiv.org/abs/0905.4749}{\tt arXiv:0905.4749}].

\bibitem{Beenakker:2009ha}
	W.~Beenakker, S.~Brensing, M.~Krämer, A.~Kulesza, E.~Laenen and I.~Niessen,
	{\it Soft-gluon resummation for squark and gluino hadroproduction},
	JHEP {\bf 0912} (2009) 041
	[\href{http://arxiv.org/abs/0909.4418}{\tt arXiv:0909.4418}].

\bibitem{Beenakker:2010nq}
	W.~Beenakker, S.~Brensing, M.~Krämer, A.~Kulesza, E.~Laenen and I.~Niessen,
	{\it Supersymmetric top and bottom squark production at hadron colliders},
	JHEP {\bf 1008} (2010) 098
	[\href{http://arxiv.org/abs/1006.4771}{\tt arXiv:1006.4771}].

\bibitem{Beenakker:2011fu}
	W.~Beenakker, S.~Brensing, M.~Krämer, A.~Kulesza, E.~Laenen, L.~Motyka and I.~Niessen,
	{\it Squark and Gluino Hadroproduction},
	Int.\ J.\ Mod.\ Phys.\ A {\bf 26} (2011) 2637
	[\href{http://arxiv.org/abs/1105.1110}{\tt arXiv:1105.1110}].

\bibitem{Beneke:2010da}
	M.~Beneke, P.~Falgari and C.~Schwinn,
	{\it Threshold resummation for pair production of coloured heavy (s)particles at hadron colliders},
	Nucl.\ Phys.\ B {\bf 842} (2011) 414
	[\href{http://arxiv.org/abs/1007.5414}{\tt arXiv:1007.5414}].

\bibitem{Falgari:2012hx}
	P.~Falgari, C.~Schwinn and C.~Wever,
	{\it NLL soft and Coulomb resummation for squark and gluino production at the LHC},
	JHEP {\bf 1206} (2012) 052
	[\href{http://arxiv.org/abs/1202.2260}{\tt arXiv:1202.2260}].

\bibitem{Langenfeld:2010vu}
	U.~Langenfeld,
	{\it Threshold Improved QCD Corrections for Stop-Antistop production at Hadron colliders},
	JHEP {\bf 1107} (2011) 052
	[\href{http://arxiv.org/abs/1011.3341}{\tt arXiv:1011.3341}].

\bibitem{Broggio:2013uba}
	A.~Broggio, A.~Ferroglia, M.~Neubert, L.~Vernazza and L.~L.~Yang,
	{\it Approximate NNLO Predictions for the Stop-Pair Production Cross Section at the LHC},
	JHEP {\bf 1307} (2013) 042
	[\href{http://arxiv.org/abs/1304.2411}{\tt arXiv:1304.2411}].

\bibitem{Beenakker:2011sf}
	W.~Beenakker, S.~Brensing, M.~Krämer, A.~Kulesza, E.~Laenen and I.~Niessen,
	{\it NNLL resummation for squark-antisquark pair production at the LHC},
	JHEP {\bf 1201} (2012) 076
	[\href{http://arxiv.org/abs/1110.2446}{\tt arXiv:1110.2446}].

\bibitem{Langenfeld:2012ti}
	U.~Langenfeld, S.~O.~Moch and T.~Pfoh,
	{\it QCD threshold corrections for gluino pair production at hadron colliders},
	JHEP {\bf 1211} (2012) 070
	[\href{http://arxiv.org/abs/1208.4281}{\tt arXiv:1208.4281}].

\bibitem{Pfoh:2013edr}
	T.~Pfoh,
	{\it Phenomenology of QCD threshold resummation for gluino pair production at NNLL},
	JHEP {\bf 1305} (2013) 044
	[JHEP {\bf 1310} (2013) 090]
	[\href{http://arxiv.org/abs/1302.7202}{\tt arXiv:1302.7202}].

\bibitem{Beenakker:2013mva}
	W.~Beenakker {\it et al.},
	{\it Towards NNLL resummation: hard matching coefficients for squark and gluino hadroproduction},
	JHEP {\bf 1310} (2013) 120
	[\href{http://arxiv.org/abs/1304.6354}{\tt arXiv:1304.6354}].

\bibitem{Beneke:2013opa}
	M.~Beneke, P.~Falgari, J.~Piclum, C.~Schwinn and C.~Wever,
	{\it Higher-order soft and Coulomb corrections to squark and gluino production at the LHC},
	PoS RADCOR {\bf 2013} (2013) 051
	[\href{http://arxiv.org/abs/1312.0837}{\tt arXiv:1312.0837}].

\bibitem{Beneke:2014wda}
	M.~Beneke, P.~Falgari, J.~Piclum, C.~Schwinn and C.~Wever,
	{\it Higher-order soft and Coulomb corrections to squark and gluino production at the LHC},
	PoS LL {\bf 2014} (2014) 060.

\bibitem{Beenakker:2014sma}
	W.~Beenakker, C.~Borschensky, M.~Krämer, A.~Kulesza, E.~Laenen, V.~Theeuwes and S.~Thewes,
	{\it NNLL resummation for squark and gluino production at the LHC},
	JHEP {\bf 1412} (2014) 023
	[\href{http://arxiv.org/abs/1404.3134}{\tt arXiv:1404.3134}].

\bibitem{Broggio:2013cia}
	A.~Broggio, A.~Ferroglia, M.~Neubert, L.~Vernazza and L.~L.~Yang,
	{\it NNLL Momentum-Space Resummation for Stop-Pair Production at the LHC},
	JHEP {\bf 1403} (2014) 066
	[\href{http://arxiv.org/abs/1312.4540}{\tt arXiv:1312.4540}].

\bibitem{Beneke:2009ye}
	M.~Beneke, M.~Czakon, P.~Falgari, A.~Mitov and C.~Schwinn,
	{\it {Threshold expansion of the $gg(q\bar{q}) \to Q\bar{Q} + X$ cross section at ${\cal O}(\alpha_s^4)$}},
	{\em Phys. Lett.} {\bf B690} (2010) 483
	[\href{http://arxiv.org/abs/0911.5166}{\tt arXiv:0911.5166}].

\bibitem{Catani:1996yz}
	S.~Catani, M.~L. Mangano, P.~Nason, and L.~Trentadue,
	{\it {The Resummation of Soft Gluons in Hadronic Collisions}},
	{\em Nucl. Phys.} {\bf B478} (1996) 273
	[\href{http://arxiv.org/abs/hep-ph/9604351}{\tt hep-ph/9604351}].

\bibitem{Agashe:2014kda}
	K.~A.~Olive {\it et al.} [Particle Data Group Collaboration],
	{\it Review of Particle Physics},
	Chin.\ Phys.\ C {\bf 38} (2014) 090001.

\bibitem{Martin:2009iq}
	A.~D.~Martin, W.~J.~Stirling, R.~S.~Thorne and G.~Watt,
	{\it Parton distributions for the LHC},
	Eur.\ Phys.\ J.\ C {\bf 63} (2009) 189
	[\href{http://arxiv.org/abs/0901.0002}{\tt arXiv:0901.0002}].

\bibitem{Vogt:2004ns}
	A.~Vogt,
	{\it Efficient evolution of unpolarized and polarized parton distributions with QCD-PEGASUS},
	Comput.\ Phys.\ Commun.\  {\bf 170} (2005) 65
	[\href{http://arxiv.org/abs/hep-ph/0408244}{\tt hep-ph/0408244}].

\bibitem{Ball:2014uwa}
	R.~D.~Ball {\it et al.} [NNPDF Collaboration],
	{\it Parton distributions for the LHC Run II},
	JHEP {\bf 1504} (2015) 040
	[\href{http://arxiv.org/abs/1410.8849}{\tt arXiv:1410.8849}].

\bibitem{Harland-Lang:2014zoa}
	L.~A.~Harland-Lang, A.~D.~Martin, P.~Motylinski and R.~S.~Thorne,
	{\it Parton distributions in the LHC era: MMHT 2014 PDFs},
	Eur.\ Phys.\ J.\ C {\bf 75} (2015) 5,  204
	doi:10.1140/epjc/s10052-015-3397-6
	[\href{http://arxiv.org/abs/1412.3989}{\tt arXiv:1412.3989}].

\bibitem{Dulat:2015mca}
	S.~Dulat {\it et al.},
	{\it The CT14 Global Analysis of Quantum Chromodynamics},
	[\href{http://arxiv.org/abs/1506.07443}{\tt arXiv:1506.07443}].

\bibitem{Beenakker:2015rna}
	W.~Beenakker, C.~Borschensky, M.~Krämer, A.~Kulesza, E.~Laenen, S.~Marzani and J.~Rojo,
	{\it NLO+NLL squark and gluino production cross-sections with threshold-improved parton distributions},
	[\href{http://arxiv.org/abs/1510.00375}{\tt arXiv:1510.00375}].

\bibitem{AbdusSalam:2011fc}
	S.~S.~AbdusSalam, B.~C.~Allanach, H.~K.~Dreiner, J.~Ellis, U.~Ellwanger, J.~Gunion, S.~Heinemeyer and M.~Kraemer {\it et al.},
	{\it Benchmark Models, Planes, Lines and Points for Future SUSY Searches at the LHC},
	Eur.\ Phys.\ J.\ C {\bf 71} (2011) 1835
	[\href{http://arxiv.org/abs/1109.3859}{\tt arXiv:1109.3859}].

\bibitem{prospino}
	W.~Beenakker, R.~Höpker, and M.~Spira, {\it\texttt{PROSPINO}: A Program for the production of supersymmetric particles in next-to-leading order QCD},
	[\href{http://arxiv.org/abs/hep-ph/9611232}{\tt hep-ph/9611232}].
	See
	\url{http://www.thphys.uni-heidelberg.de/~plehn/index.php?show=prospino} or
	\url{http://tiger.web.psi.ch/prospino/}, 1996.

\bibitem{nllfast}
	Publicly available from \url{http://pauli.uni-muenster.de/~akule_01/nllwiki/index.php/NLL-fast}.

\end{thebibliography}
\providecommand{\href}[2]{#2}\begingroup\raggedright\endgroup

\end{document}